\documentclass[12pt]{article}
\usepackage{graphicx}
\usepackage[figuresright]{rotating}
\newcommand{\beq}{\begin{equation}}
\newcommand{\eeq}{\end{equation}}
\newcommand{\be}{\begin{eqnarray}}
\newcommand{\ee}{\end{eqnarray}}
\newcommand{\ba}{\begin{array}}
\newcommand{\ea}{\end{array}}

\newcommand{\displayfrac}[2]{\displaystyle{\frac{#1}{#2}}}

\newcommand{\nablaslash}{\nabla\hspace{-.65em}/\hspace{.3em}}

\newcommand{\Sslash}{S\hspace{-.65em}/\hspace{.3em}}
\newcommand{\partialslash}{\partial\hspace{-.5em}/\hspace{.15em}}

\newcommand{\Mn}{ M_{\mbox{\tiny N}}}

\renewcommand{\theequation}{\arabic{section}.\arabic{equation}}
\begin{document}
%
%
\rightline{RUB-TPII-15/00}
\rightline{hep-ph/0101300}
\vspace{.3cm}
\begin{center}
\begin{large}
{\bf Transversity distributions in the nucleon in the large--$N_c$ limit} 
\end{large}
\\[1cm]
{\bf P. Schweitzer$^{a,\,\rm 1}$}, 
{\bf D. Urbano$^{b,\,\rm 2}$},
{\bf M.V.\ Polyakov$^{a,\,c,\,\rm 3}$},\\
{\bf C.\ Weiss$^{d,\, \rm 4}$},
{\bf P.V.\ Pobylitsa$^{a,\,c,\,\rm 5}$}, 
{\bf K. Goeke$^{a,\,\rm 6}$}
\\[0.2 cm]
{\footnotesize \it 
$^a$ 
Institut f\"ur Theoretische Physik II,
Ruhr--Universit\"at Bochum, D--44780 Bochum, Germany \\
$^b$ 
Faculdade de Engenharia da Universidade do Porto, 
4000 Porto, Portugal\\
$^c$ 
Petersburg Nuclear Physics Institute, Gatchina,
St.\ Petersburg 188350, Russia \\
$^d$
Institut f\"ur Theoretische Physik, Universit\"at Regensburg,
D-93053 Regensburg, Germany\\} 

\end{center}
\vspace{1.5cm}
\begin{abstract}
\noindent
We compute the quark and antiquark transversity distributions
in the nucleon at a low normalization  point ($\mu\!\approx\!600\,{\rm MeV}$)
in the large--$N_c$ limit, where the nucleon can be described 
as a soliton of an effective chiral theory (chiral quark--soliton
model). The flavor--nonsinglet distributions, $\delta u(x)\!-\!\delta d(x)$ 
and $\delta\bar u(x)\!-\!\delta\bar d(x)$, appear in leading order of
the $1/N_c$--expansion, while the flavor--singlet distributions,
$\delta u(x)\!+\!\delta d(x)$ and $\delta\bar u(x)\!+\!\delta\bar d(x)$,
are non-zero only in next--to--leading order. 
The transversity quark and antiquark distributions are found to be 
significantly different from the longitudinally polarized distributions 
$\Delta u(x)\!\pm\!\Delta d(x)$ and $\Delta\bar{u}(x)\!\pm\!\Delta\bar{d}(x)$,
respectively, in contrast to the prediction of the naive non-relativistic 
quark model. 
We show that this affects the predictions for the spin asymmetries in 
Drell--Yan pair production in transversely polarized pp and p$\bar{\rm p}$ 
collisions.
\end{abstract}
\vspace{1cm}
PACS: 13.60.Hb, 12.38.Lg, 11.15.Kc, 12.39.Ki \\
Keywords: \parbox[t]{13cm}{polarized parton distributions, transversity,
$1/N_c$--expansion, chiral quark--soliton model of the nucleon}
\vfill
\rule{5cm}{.15mm}
\\
\noindent
{\footnotesize $^{\rm 1}$ E-mail: peterw@tp2.ruhr-uni-bochum.de} \\
{\footnotesize $^{\rm 2}$ E-mail: dianau@tp2.ruhr-uni-bochum.de} \\
{\footnotesize $^{\rm 3}$ E-mail: maximp@tp2.ruhr-uni-bochum.de} \\
{\footnotesize $^{\rm 4}$ E-mail: christian.weiss@physik.uni-regensburg.de}\\
{\footnotesize $^{\rm 5}$ E-mail: pavelp@tp2.ruhr-uni-bochum.de} \\
{\footnotesize $^{\rm 6}$ E-mail: goeke@tp2.ruhr-uni-bochum.de} 
%
%
\newpage
\tableofcontents
\newpage
\section{Introduction}
\setcounter{equation}{0}
A central property of QCD is the possibility to factorize the cross sections 
for some hard scattering processes involving hadrons into the cross section 
for the partonic subprocess involving quarks and gluons, calculable within 
perturbative QCD, and certain functions describing the transition from 
hadrons to quark and gluon degrees of freedom. 
Of the latter only the scale dependence can be predicted from perturbative 
QCD; the value of the functions themselves at a low scale can only be inferred
from experiment, or be calculated using non-perturbative methods. 
In inclusive hard scattering off the nucleon [deep--inelastic scattering 
(DIS)], semi--inclusive particle production, and Drell--Yan pair production, 
the relevant characteristics of the nucleon are the so-called quark and 
antiquark distributions in the nucleon -- 
functions describing the emission in the collinear direction and subsequent 
absorption by the nucleon of a quark/antiquark carrying a fraction $x$ of the 
nucleon momentum. 
At twist--2 level, helicity conservation allows for three different types of 
such functions \cite{Jaffe97}. 
The usual unpolarized and longitudinally polarized distributions, 
$q(x), \, \bar q(x)$ and $\Delta q(x), \, \Delta \bar q(x)$, correspond to 
probabilities for emitting and absorbing a quark or antiquark of same 
helicity, and can be interpreted as the sum and difference of probabilities 
to find a quark/antiquark with longitudinal polarization parallel or 
antiparallel to that of the nucleon. 
The third kind of functions, $\delta q (x), \, \delta \bar q (x)$, describe 
the emission and absorption of quarks/antiquarks with different helicities 
\cite{RS79}. 
They can be interpreted as the probability to find a transversely polarized 
quark/antiquark in a transversely polarized nucleon and are therefore
referred to as transversity distributions. 
\par
The crucial difference between the transversity and the ``usual'' 
distributions lies in their chirality properties. 
The unpolarized and longitudinally polarized distributions parameterize 
the chirally even parts of the quark density matrix in the nucleon 
(Dirac structures $\gamma^+$ and $\gamma^+ \gamma_5$, respectively). 
They can therefore be measured in DIS at leading--twist level, where the 
chirality of the quarks is preserved by the hard scattering process.
These distributions are by now well--known,
with most of the information coming from QCD fits to DIS 
data \cite{GRSV96,GRV95,MRS98,CTEQ99}.
In contrast, the transversity distributions describe the chirally odd part of 
the quark density matrix (Dirac structure $\sigma^{+\perp} \gamma_5$). 
Consequently, they can be measured only in hard processes where they enter 
together with other chirally--odd objects, chirally--odd distribution or 
fragmentation functions. 
A variety of such processes -- both hadron--hadron and lepton--hadron induced 
-- are currently being considered; see Ref.\cite{Jaffe97} for a review.
In hadron--hadron collisions the transversity distribution can be measured
in Drell--Yan pair production with transversely polarized protons, where one 
combines the transversity quark and antiquark distributions in the two 
protons \cite{RS79,JJ91}, or in polarized jet production 
\cite{Ji92,JaffeSaito96}. In lepton--hadron scattering transversity
is in principle accessible through semi--inclusive particle production, where 
the transversity distribution in the target is combined with a 
chiral--odd fragmentation function. In single--particle production one can 
measure the left--right asymmetries in the fragmentation of 
a transversely polarized quark, which are produced by $T$--odd fragmentation 
functions (Collins effect) 
\cite{Collins,Kotzinian:1997wt,Boer98,Efremov:1999mv,Korotkov:1999jx,
Kotzinian:1999dy,Boglione:2000jk}. 
In such processes there is usually a competing contribution involving 
a higher--twist distribution function in the target and a chiral--even 
fragmentation function, which makes extraction of the transversity 
distribution difficult. 
First estimates of the transversity distributions have been reported 
\cite{Efremov:1999mv,Diana1} based on the azimuthal asymmetries measured by 
the HERMES \cite{Avakian:1999rr} and SMC \cite{Bravar:1999rq} experiments. 
Alternatively, one can consider semi-inclusive production of two pions,
where a $T$--odd structure in the fragmentation function appears
due to the possibility of interference between S-- and P--waves 
in the hadronic final state \cite{JJT97}. It has also been
suggested to use so--called ``handedness'' correlations in multi-particle
production to access the transversity distributions \cite{Efremov:1999mv}.
\par
The transversity distributions have many interesting theoretical aspects, 
related to general properties of QCD as well as to the structure of 
the nucleon. 
In leading order (LO), certain inequalities have been derived by Soffer 
\cite{Soffer95}, giving an upper bound on the transversity distribution in 
terms of the longitudinally polarized and unpolarized quark distributions; 
see also Refs.\cite{GJJ95}. 
Their $Q^2$--evolution and generalization to NLO have been discussed in 
Ref.\cite{Bourrely:1998bx}
Also, it has long been noted that in the non-relativistic quark model 
(no spin--orbit interaction) the transversity distributions are identical 
to the longitudinally polarized ones.
In this sense, any measurements of the transversity distribution would have 
implications for our understanding of the relativistic structure of the 
nucleon.
\par
Recently a method has been formulated to calculate the
quark-- and antiquark distributions in the nucleon at a low
normalization point in the large--$N_c$ limit \cite{DPPPW96,DPPPW97}. 
In this limit
the nucleon can be described as a soliton of an effective low--energy
theory based on the dynamical breaking of chiral symmetry in 
QCD (chiral quark--soliton model) \cite{DPP88}.
It has been shown that this fully field--theoretic description of the 
nucleon preserves all qualitative properties of the quark-- and 
antiquark distributions known from QCD, such as positivity conditions, 
sum rules {\it etc.}. The unpolarized distributions have 
been computed in Refs.\cite{DPPPW96,DPPPW97,PPGWW98}; the results are 
in good agreement with the well--known parameterizations of the parton 
distribution functions at a low input scale \cite{GRV95}. In particular,
this approach describes well the observed violation of the Gottfried
sum rule and the flavor asymmetry of the unpolarized antiquark 
distribution \cite{WK98,PPGWW98,Dressler98,Wakamatsu:2000nj,Dressler:2000zg}. 
The longitudinally polarized
flavor--singlet distribution calculated in this approach \cite{WK98}
is also in good agreement with the parameterizations of the polarized
DIS data \cite{GRSV96}. Interestingly, the large--$N_c$ approach
predicts a large flavor asymmetry of the polarized antiquark
distribution \cite{DPPPW96,DPPPW97,Dressler98,Wakamatsu:2000nj},
which could produce observable effects {\it e.g.}\ 
in semi--inclusive spin asymmetries \cite{Dressler:2000zg}, 
and in Drell-Yan double-spin asymmetries with polarized protons
\cite{Dressler:1999zv}.
Also, the flavor--nonsinglet transversity distribution has been computed in 
Ref.\cite{PP96}; the isosinglet one in Ref.\cite{WK98,Wakamatsu:2000fd}. 
We also note that the chiral quark--soliton model has been successfully 
applied to calculate skewed (non-diagonal) parton 
distributions \cite{PPPBGW97}. The model also describes well 
a large number of hadronic observables of the nucleon as well
as the other octet and decuplet baryons, such as magnetic 
moments, electromagnetic and weak form factors, {\it etc.};
see Refs.\cite{Review,Alkofer:1996ph} for a review.
\par
In this paper we report about a comprehensive investigation of the 
quark-- and antiquark transversity distributions in the
large--$N_c$ approach of Refs.\cite{DPPPW96,DPPPW97}. We
calculate both the flavor--nonsinglet distribution, which appears
in leading order of the $1/N_c$ expansion, and the flavor--singlet
one, which is non-zero only in next-to-leading order. In Ref.\cite{PP96}
an approximation was used in dealing with the contribution of the Dirac 
continuum of quark states in the soliton; in this paper we perform 
a numerical calculation of the distribution functions without further
approximations. This is particularly important for the
antiquark transversity distribution. 
\par
The effective chiral theory used to describe the nucleon 
is non-renormalizable and defined with an explicit ultraviolet cutoff.
An important problem when calculating parton distributions in this
approach is to ensure that none of their essential properties such
as positivity, sum rules {\it etc.}\ are violated
by the ultraviolet regularization. For the unpolarized and polarized
distributions this problem has been studied extensively in 
Refs.\cite{DPPPW96,DPPPW97}. Here we use the methods developed there
to investigate the role of the UV cutoff on the transversity 
distributions. It turns out that these distributions are UV finite
and can eventually be computed without any cutoff. 
Still, in the actual calculation of the distributions one has to introduce 
a cutoff during the intermediate stages, and one must make sure that the 
finite result in the end does not depend on this intermediate
regularization procedure.        
We explicitly address this problem here. In particular, we shall 
describe an interesting anomaly--type phenomenon observed in the 
calculation of the transversity distributions, which can be stated as 
the non-commutativity of the limits of infinite UV cutoff and
the chiral limit (vanishing pion mass). A thorough understanding of this
phenomenon is a prerequisite for a reliable calculation of the
transversity distributions in the effective chiral theory.
\par
The plan of this paper is as follows. In Section~\ref{sec_QCD} we
summarize the definition and important properties of the transversity 
distributions and the tensor charges within QCD. 
Section~\ref{sec_effective} gives a detailed exposition of the
method used to calculate the transversity quark-- and antiquark 
distributions at a low normalization point. After a brief description
of the effective low--energy theory and the chiral quark--soliton model
of the nucleon in Section~\ref{subsec_model} we discuss in 
Section~\ref{subsec_N_c} the $1/N_c$ expansion of the transversity 
distributions and the tensor charges. The calculation of the 
transversity distributions --- both isovector and isoscalar -- in the 
chiral quark--soliton model is described in Section~\ref{subsec_calculation}. 
In particular, in this section we derive expressions for these 
distributions in the form of sums over contributions of quark 
single--particle levels in the soliton field which are used in the 
numerical calculations. Section~\ref{subsec_regularization} deals
with the important issue of ultraviolet regularization. The 
dependence of the transversity distributions on the ultraviolet cutoff
is studied using analytic methods (gradient expansion). In 
Section~\ref{subsec_anomaly} we describe the anomaly--type phenomenon
in the tensor charge, which is both of general theoretical interest 
as well as of importance in the numerical calculations. The nature of this
phenomenon is clarified, and an explicit expression for the anomalous 
difference of the sums over contributions of occupied and non-occupied 
quark levels is derived. The corresponding difference for the 
full $x$--dependent distribution is given in Appendix~\ref{app_anomaly}.
In Section~\ref{sec_results} we describe the results for the 
transversity distributions at the low scale. In 
Section~\ref{subsec_transverse_vs_longitudinal} we compare the 
results for the transversity distributions to those for the 
longitudinally polarized ones \cite{Goeke:2000ym,Goeke:2000wv} 
at a qualitative level,
considering various limiting cases of the chiral quark--soliton model
which correspond to the non-relativistic quark model or the
Skyrme model. Numerical results are presented in 
Section~\ref{subsec_numerical}, where also a brief description of the
numerical method used to evaluate the sums over quark levels is given.
In Section~\ref{subsec_inequalities} we demonstrate that the calculated
transversity distributions satisfy the recently derived 
``large--$N_c$ versions'' of positivity and Soffer inequalities
\cite{Pobylitsa:2000tt}, which impose stronger constraints on the 
distributions than the ``usual QCD'' positivity and Soffer inequalities.
Finally, in Section~\ref{sec_asymmetries} we use our results
for the transversity distributions to make predictions for the
double spin asymmetries in  Drell--Yan pair production in 
polarized $pp$ and $p\bar p$ scattering. In particular, we compare
the asymmetries calculated from our model distributions with those
obtained by assuming $\delta q (x) \equiv \Delta q (x)$, an approximation
frequently made in the literature. We find that our model distributions
result in significant deviations from what is obtained with that
approximation, which makes us hope that RHIC measurements may be able
to discriminate between the two predictions.
\par
The transversity distributions at a low normalization point have been 
studied in a variety of other approaches. Model calculations of the
distributions have been performed within the bag model \cite{JJ91}
and a constituent quark model \cite{Scopetta:1998qg}. Furthermore,
these distributions have been studied in the chiral quark--soliton 
model \cite{WK98,PP96} and a closely related description of the
nucleon as a chiral soliton of the Nambu--Jona Lasinio 
model \cite{Gamberg:1998vg}. 
The transversity distributions have also been estimated
using QCD sum rule techniques \cite{Ioffe:1995aa}. 
For the tensor charges of the
nucleon estimates from lattice 
simulations \cite{Aoki:1997pi,Kuramashi:1998fi}
and QCD sum rules \cite{HeJi96} are available.
The tensor charges have also been calculated within the
the chiral quark--soliton model \cite{Kim96} (the relation of these results
to the present calculation is discussed in 
Section~\ref{subsec_numerical}).
The evolution (scale dependence) of the transversity distributions
has been studied by a number of groups, see
Refs.\cite{Artru:1990zv,Vogelsang:1998ak,Hayashigaki:1997dn}.
\section{Transversity distributions in QCD}
\setcounter{equation}{0}
\label{sec_QCD}
In QCD, parton distributions are defined
as expectation values of certain twist--2 light--ray operators
in the nucleon state. The helicity decomposition of the 
matrix element of a general quark bilinear in a polarized nucleon
(``spin density matrix'') contains three twist--2 structures, which 
can be identified as the unpolarized, longitudinally polarized,
and transversity distribution \cite{Jaffe97}. The explicit expression for 
the transversity distribution is
\beq
\delta q_f (x) \;\; = \;\; \int\frac{dz^-}{4\pi} \; e^{i x P^+ z^-}
\langle N (P), S_\perp |\; \bar\psi_f(0) \gamma^+
\gamma_5 \Sslash_\perp \, [0, z] \, \psi_f (z) \; | N(P), S_\perp 
\rangle_{z^+ = z_\perp = 0} ,
\label{delta_q_def}
\eeq
where $\psi_f$ is the quark field of flavor $f$, $z$ a light--like 
distance ($z^\mu z_\mu \!=\! 0$), and the
light--cone coordinates are defined 
as\footnote{It is convenient here to define the light--cone
vector components using the $1$-- rather than the $3$--direction. 
With this choice the direction of transverse polarization of the 
nucleon can be chosen as the $3$--direction, which will make
the expressions in Subsection \ref{subsec_calculation} look more
conventional.}
\be
z^\pm &=& \frac{z^0 \pm z^1}{\sqrt{2}}, 
\hspace{2cm}
{\bf z}^\perp \;\; =  \;\; (z^2, z^3) .
\ee
Furthermore, $[0, z]$ denotes the link operator
(path--ordered exponential of the gauge field) required by
gauge invariance. The nucleon state in Eq.(\ref{delta_q_def})
is transversely polarized
($S_\perp^2 = -1, \; P_\mu S_\perp^\mu = 0$), and 
$\Sslash_\perp \equiv \gamma_\mu S_\perp^\mu$. In the nucleon rest frame, 
the components of the polarization vector are given by 
\be
S_\perp^\mu &=& (0,\, 0,\, 0,\, \lambda)
\ee
where $\lambda\!=\!2S_3\!=\!\pm 1$ and
$S_3$ is the spin projection on the $3$--axis.
In this frame Eq.(\ref{delta_q_def}) reduces to
\be
\delta q_f (x) =
- (2 S_3) \!\int\!\frac{d z^0}{4\pi}\,e^{i x M_N z^0}
\langle N, S_3 | \, \bar\psi_f (0) (\gamma^0 \!+\!\gamma^1)
\gamma_5 \gamma^3 \, [0, z] \, \psi_f (z) \, |N, S_3  
\rangle\biggl|_{\!\!\renewcommand{\arraystretch}{0.3}
         \begin{array}{l}\scriptscriptstyle {z^1\,   =-z^0} \\
                         \scriptscriptstyle {z_\perp = 0} \end{array}} . 
\label{delta_q_rf}
\ee
\par
The operator in Eq.(\ref{delta_q_def}) is $C$--odd, so the 
corresponding antiquark distribution is obtained from
the expression on the R.H.S.\ of Eq.(\ref{delta_q_def}) by
\be
\delta \bar q_f (x) &=& -\delta q_f(-x).
\label{delta_qbar_def}
\ee 
Furthermore, the light--ray operator in Eq.(\ref{delta_q_def})
is scale dependent, so that the 
distributions depend on the normalization point. 
This dependence can be described by evolution equations
for the transversity distributions, which have been
studied in Refs.\cite{Artru:1990zv,Vogelsang:1998ak,Hayashigaki:1997dn}.
\par
The normalization integral for the distribution Eqs.(\ref{delta_q_def}) 
and (\ref{delta_qbar_def}) is given by the so-called tensor charge
of the nucleon,
\be
\delta q_f &\equiv& \int_0^1 dx \, 
\left[ \delta q_f (x) - \delta \bar q_f (x) \right]
\ee
which is defined as the matrix element of the local 
pseudo--tensor operator,
\be
\langle N(P), S | \bar\psi  (-i) \sigma^{\mu\nu} \gamma_5
\psi  | N(P), S \rangle &=& (P^\mu S^\nu - P^\nu S^\mu ) \;
\delta q_f .
\label{tensor_def}
\ee
In particular, in the nucleon rest frame it is given by
\be
\langle N, S_3 | \bar\psi  \gamma^0 \gamma^3 \gamma_5
\psi  | N, S_3 \rangle &=& 2 M_N (2 S_3) \delta q_f .
\label{tensor_rf}
\ee
\section{Transversity distributions from the effective chiral theory}
\setcounter{equation}{0}
\label{sec_effective}
\subsection{The nucleon as a chiral soliton}
\label{subsec_model}
It is generally believed that in the large--$N_c$ limit QCD
becomes equivalent to a theory of mesons, with baryons emerging
as solitonic excitations \cite{Witten83}. While this connection
alone has many interesting qualitative implications, it is not 
known at present how to derive this effective theory from 
QCD in full generality. However, quantitative calculations 
within the large--$N_c$ ideology can be done in certain limiting 
cases where the dynamics of QCD simplifies. It is known that
at energies far below the mass of mesonic resonances (say, the
rho meson) the dynamics of strong interactions is
governed by the spontaneous breaking of chiral symmetry.
In fact, in the long wavelength limit the dynamics is completely
described by the chiral Lagrangian containing only the Goldstone
boson (pion) field to some given order in derivatives. To describe 
the nucleon as a chiral soliton, however, one needs an effective 
theory valid in a wider region of momenta. Such a theory has been 
derived within the framework of the instanton description of the 
QCD vacuum, which provides a ``microscopic'' picture of the 
dynamical breaking of chiral symmetry in QCD \cite{DP86}.
In this approach the effective action for the pion field
is obtained in the form of an integral over quark fields,
which have obtained a dynamical mass in the spontaneous breaking
of chiral symmetry, and which interact with the pion field in a 
chirally invariant way \cite{DP86,DE}:
\be
\exp\left( i S_{\rm eff}[U (x)] \right) &=&
\int D\psi D\bar\psi \; \exp\left[ i\int d^4 x\,
\bar\psi(i\partialslash - M U^{\gamma_5})\psi\right] .
\label{effective_action}
\ee
Here, $\bar\psi , \psi$ are the fermion fields, $M$ is the dynamical 
quark mass, and the Goldstone boson field is parameterized as a unitary 
matrix, $U(x)$, with
\be
U^{\gamma_5}(x) &=& \frac{1+\gamma_5}2 U (x)
+ \frac{1-\gamma_5}2 U^\dagger (x) .
\label{U_gamma5}
\ee
In the long--wavelength limit, when expanding in derivatives of the 
pion field, the effective action Eq.(\ref{effective_action}) reproduces the
chiral Lagrangian with correct coefficients, including the
Gasser--Leutwyler $O(p^4 )$ terms and the Wess--Zumino term. 
However, the validity of the theory defined by Eq.(\ref{effective_action}) 
is not restricted to the long--wavelength limit.
\par
In the effective theory derived from the instanton vacuum the
dynamical quark mass in Eq.(\ref{effective_action}) is 
momentum--dependent and drops to zero for momenta of the 
order of the inverse average instanton size, 
$\bar\rho^{-1} \simeq 600 \, {\rm MeV}$. This provides for a
natural ultraviolet cutoff of the effective theory. 
In the calculation of quark distribution functions here
we take the dynamical quark mass in Eq.(\ref{effective_action})
to be constant and simulate the ultraviolet cutoff implied
by the instanton vacuum by applying an external ultraviolet 
regularization to divergent quark loops. We shall show in 
detail that this is a legitimate approximation for the
quantities considered here.
\par
The effective action allows to compute hadronic correlation
functions at low energies in the large--$N_c$ limit.
In particular, the nucleon in the effective theory
at large--$N_c$ limit characterized by a classical pion
field which binds the quarks (chiral quark--soliton model) \cite{DPP88}.
In the nucleon rest frame the classical pion field is of ``hedgehog'' 
form,
\be
U_{\rm cl} ({\bf x}) &=& \exp\left[ i \frac{x^a \tau^a}{r} P(r) \right] ,
\hspace{2cm} r \;\; = \;\; |{\bf x}|,
\label{hedge}
\ee
where $P(r)$ is called the profile function, with $P(0) = -\pi$ and 
$P(r) \rightarrow 0$ for $r\rightarrow\infty$. The quarks are described by
single--particle wave functions, which are the solutions of the Dirac 
equation in the background pion field, Eq.(\ref{hedge}):
\be
H(U_{\rm cl} ) |n\rangle &=& E_n |n\rangle ,
\label{eigen_n}
\ee
where
\be
H(U_{\rm cl}) &=& - i\gamma^0\gamma^k \partial_k + M\gamma^0 U_{\rm cl}^{\gamma_5} 
\label{H}
\ee
is the single--particle Dirac Hamiltonian in the classical 
background pion field. The spectrum of $H(U_{\rm cl})$ includes a 
discrete bound--state level, whose
energy is denoted by $E_{\rm lev}$, as well as the positive and negative
Dirac continuum, polarized by the presence of the pion field.  The soliton
profile, $P(r)$, is determined by minimizing the static energy of the pion
field, which is given by the sum of the energy of the bound--state level
and the aggregate energy of the negative Dirac continuum, the energy of the
free Dirac continuum $(U = 1)$ subtracted \cite{DPP88},
\be
E_{\rm tot}[U_{\rm cl}] &=& N_c \left[ 
\sum_{\scriptstyle n \atop \scriptstyle {\rm occup.}} E_n 
- \sum_{\scriptstyle n \atop \scriptstyle {\rm occup.}} E_n^{(0)} 
\right]
\nonumber \\
&=& N_c E_{\rm lev} \; + \; 
N_c \sum_{\scriptstyle n \atop \scriptstyle {\rm neg. cont.}} 
( E_n - E_n^{(0)} ) ,
\label{E_tot}
\ee
and in the leading order of the $1/N_c$--expansion the nucleon mass is
given simply by the value of the energy at the minimum,
\be
M_N &=& E_{\rm tot} [U_{\rm cl}] .
\label{M_N_static}
\ee
The expression for the energy of the pion field, Eq.(\ref{E_tot}), contains
a logarithmic ultraviolet divergence due to the contribution of the Dirac
continuum and requires ultraviolet regularization (see below).  
\par
In higher order of the $1/N_c$--expansion one must take into account the
fluctuations of the pion field about its saddle--point value. A special role
is played by the zero modes of the pion field. The minimum of the energy,
Eq.(\ref{E_tot}), is degenerate with respect to translations of the soliton
field in space, and to rotations in ordinary and isospin space [for the
hedgehog field, Eq.(\ref{hedge}), the two types of rotations are
equivalent]. Quantizing these zero modes modes gives rise to nucleon 
states with definite momentum and spin/isospin quantum 
numbers \cite{ANW,DPP88}. This is done by subjecting the hedgehog field,
Eq.(\ref{hedge}), to time--dependent translations and flavor rotations,
\be
U_{\rm cl} ({\bf x}) &\rightarrow& R(t) \, 
U_{\rm cl}({\bf x} - {\bf X}(t)) \, 
R^\dagger (t) ,
\label{rotation_U}
\ee
where $R(t)$ is an $SU(2)$ rotation matrix,
and computing the functional integral over the collective
coordinates within the $1/N_c$--expansion. The functional integral 
over translations, ${\bf X}(t)$, can be reduced to a Hamiltonian system 
describing the free motion of the soliton center--of--mass, with
mass $M_N$, Eq.(\ref{M_N_static}), whose eigenfunctions are plane waves 
with given three--momentum, $\exp ( i \, {\bf P}\!\cdot\!{\bf X})$. [The 
$O(1/N_c)$ contribution of the collective translations to the nucleon 
mass can be neglected in the applications considered here.]
In the functional integral over collective rotations, $R(t)$, the 
rotational energy is given by
\begin{eqnarray}
E_{\mbox{\tiny rotating soliton}} -
E_{\mbox{\tiny static soliton}}
=  I\, \mbox{tr} \left[ \Omega (t) \Omega (t) \right] \; + \;\dots \; , 
\end{eqnarray}
where
\begin{eqnarray}
\Omega (t) \; \equiv \;  \Omega_a (t) \frac{\tau^a}{2}
\; = \; -i R^\dagger(t) \dot{R}(t)\; &&
\label{omega_def}
\end{eqnarray}
is the angular velocity, and $I$ is the moment of inertia of the
soliton, which is given by a double sum over quark single--particle 
levels in the background pion field,
\be
I &=& \frac{N_c}{6} \sum_{\scriptstyle n \atop \scriptstyle {\rm occup.}}
\sum_{\scriptstyle m \atop \scriptstyle {\rm non-occup.}}
\frac{\langle n | \tau^a | m \rangle \langle m | \tau^a | n \rangle}
{E_m - E_n} .
\label{I}
\ee
Here the sum over $n$ runs over all occupied states, {\it i.e.}, the
discrete level and the negative Dirac continuum, the sum over $m$ over all
non-occupied states, {\it i.e.}, the positive Dirac continuum.
(The ultraviolet regularization of this quantity will be discussed 
below.) It is important that the moment of inertia 
is of order $N_c$, so the typical angular velocities are
\be
\Omega (t) &\sim& \frac{1}{N_c}
\ee
and one can compute the functional integral by expanding in powers 
of the angular velocity. To leading order in $1/N_c$ the collective 
motion is described by a Hamiltonian
\be
H_{\rm rot} &=& \frac{S_a^2}{2 I} \;\; = \;\; \frac{T_a^2}{2 I} ,
\label{H_rotator}
\ee
where $S_a$ and $T_a$ are the right and left angular momenta, and the
Hamiltonian Eq.(\ref{H_rotator}) has been obtained by the ``quantization
rule''
\be
\Omega_a &\rightarrow& \frac{S_a}{I} .
\label{quantization_rule}
\ee
This Hamiltonian describes a spherical top in spin/isospin space, 
subject to the constraint $S^2 = T^2$, which is a
consequence of the ``hedgehog'' symmetry of the static pion field,
Eq.(\ref{hedge}). Its eigenfunctions, classified by $S^2 = T^2, S_3$ and
$T_3$ are given by the Wigner finite--rotation matrices \cite{DPP88},
\be
\phi^{S=T}_{S_3 T_3}(R) &=&
\sqrt{2S+1} (-1)^{T+T_3} D^{S=T}_{-T_3,S_3}(R) .
\label{Wigner}
\ee
The four nucleon states have $S = T = 1/2$, with $S_3, T_3 = \pm 1/2$,
while for $S = T = 3/2$ one obtains the 16 states of the $\Delta$
resonance.  The rotational energy, $S(S + 1)/(2 I)$, gives a
$1/N_c$--correction to the nucleon mass, which should be added to
Eq.(\ref{M_N_static}).  In particular, the nucleon--$\Delta$ mass splitting
is given by
\be
M_\Delta - M_N &=& \frac{3}{2 I} .
\ee
\par
The saddle--point solution, Eq.(\ref{hedge}), and the collective
quantization procedure outlined above, not only give rise to nucleon
states of definite momentum and spin/isospin quantum numbers in the 
effective chiral theory defined by Eq.(\ref{effective_action}), they also
imply a prescription for the calculation of matrix elements
of arbitrary composite quark operators between nucleon states.
For a review of the applications of this model to baryon observables
such as masses, form factors, {\it etc.}\ we refer to Ref.\cite{Review}.
We shall apply this approach to calculate the 
nucleon's transversity distributions in 
Subsection \ref{subsec_calculation}.
\subsection{The transversity distributions in the
large--$N_c$ limit}
\label{subsec_N_c}
Before embarking on the calculation of the transversity 
distributions in the chiral quark soliton model it is useful
to establish the large--$N_c$ behavior of these distributions
on general grounds \cite{DPPPW96,PP96}. Standard $N_c$ counting
tells us that at large $N_c$ the tensor charges of the 
nucleon scale as
\be
\delta u - \delta d &\sim& N_c, \hspace{2cm}
\delta u + \delta d \;\; \sim \;\; 1,
\label{tensor_large_N}
\ee
{\it i.e.}, the isovector matrix element is leading relative
to the isosinglet one. This behavior is analogous to that
of the axial charges, which scale as
\be
g_A^{(3)} &=& \Delta u - \Delta d \;\; \sim \;\; N_c, \hspace{2cm}
g_A^{(0)} \;\; = \;\; \Delta u + \Delta d \;\; \sim \;\; 1.
\ee
Combining Eq.(\ref{tensor_large_N}) with the fact that in 
the large--$N_c$ limit the parton distributions are concentrated
at values of $x$ of the order of $1/N_c$ one obtains that
the transversity distributions at large $N_c$ scale as
\be 
\delta u(x) - \delta d(x), \;\; \delta \bar u(x) - \delta \bar d(x)
&\sim& N_c^2 \,\, f(N_c x),
\label{Nc_large}
\\
\delta u(x) + \delta d(x), \;\; \delta \bar u(x) + \delta \bar d(x)
&\sim& N_c \,\, f(N_c x) ,
\label{Nc_small}
\ee
where $f(y)$ is a stable function in the large $N_c$--limit, which
depends on the particular distribution considered.
\subsection{Calculation of the transversity distributions}
\label{subsec_calculation}
We now outline the calculation of the transversity quark and antiquark 
distributions in the chiral quark soliton model. The methods
for computing parton distributions at a low normalization point in 
this approach have been developed in Refs.\cite{DPPPW96,DPPPW97,PPGWW98},
and we refer to these papers for all general methodological questions.
\par
When computing quark/antiquark distribution in the effective chiral
theory it is assumed that the normalization point of the distributions
is of the order of the ultraviolet cutoff of the effective theory,
i.e. of ${\cal O}(600\,{\rm MeV})$. 
At this scale the QCD twist--2 quark operators
can be identified with the corresponding operators in the effective 
theory. The gluon distributions are zero at this level of 
approximation \cite{DPPPW96}. A more explicit justification for this
procedure is provided by the instanton picture of the QCD vacuum,
which allows to derive the low--energy effective theory. 
It was shown that in leading order of the packing fraction of the
instanton medium the quark and antiquark distributions saturate
the nucleon momentum and spin sum rule, while the gluon distributions
are zero \cite{DPPPW96,BPW97}. The calculations of parton distributions
in Refs.\cite{DPPPW96,DPPPW97,WG97,PPGWW98,Goeke:2000ym,Goeke:2000wv} were 
performed at this level of approximation. In the case of the transversity 
distributions, since there is no transversity gluon distribution, 
one may expect this ``quarks--antiquarks only'' approximation to give even 
better results than in the unpolarized and longitudinally polarized case.
\par
Having expressed the QCD operators in 
Eqs.(\ref{delta_q_rf}, \ref{delta_qbar_def}) in terms of the
quark fields of the effective theory, we can now compute their
nucleon matrix elements within the $1/N_c$--expansion, using
standard techniques. It is convenient to work in the nucleon
rest frame, where the classical pion field describing the
nucleon (up to collective translations and rotations) is given 
by Eq.(\ref{hedge}); in this frame the matrix elements defining
the transversity distributions take the form 
Eqs.(\ref{delta_q_rf}, \ref{delta_qbar_def}).
Matrix elements of quark bilinear operators such as 
Eqs.(\ref{delta_q_rf}, \ref{delta_qbar_def}) can be 
reduced to those of time--ordered products of quark fields,
which can be calculated with the help of the Feynman Green function 
of the quarks in the background pion field,
\be
G_F (y^0, {\bf y}; x^0, {\bf x})
&=&
\langle y^0, {\bf y} | \left[ i\partial_t - H(U) \right]^{-1}
| x^0, {\bf x} \rangle .
\label{G_F}
\ee
Here the saddle--point pion field is the slowly rotating hedgehog field,
Eq.(\ref{rotation_U}). For this ansatz the Green function, Eq.(\ref{G_F}),
takes the form
\be
\left[
i\partial_t - H(U) \right]^{-1} &=&
R(t) \; [i\partial_t - H(U_{\rm cl}) - \Omega (t) ]^{-1} \; R^\dagger (t) ,
\label{rotH}
\ee
where $\Omega (t)$ is the angular velocity, Eq.(\ref{omega_def}).
Performing the functional integral over collective coordinates
of the saddle--point field as described in 
Subsection \ref{subsec_model}, projecting on nucleon states
with definite momentum and spin/isospin quantum numbers
one obtains the following ``master formula'' for the
expectation value of a color--singlet time--ordered quark 
bilinear operator in the nucleon:
\be
\lefteqn{
\langle {\bf P} = 0, \, S=T,\, S_3, \, T_3 | \,
{\rm T} \left\{ \psi^\dagger(x) \Gamma 
\psi(y) \right\} | {\bf P} = 0, \, S=T,\, S_3,\, T_3 \rangle }
\nonumber \\
&=& 2 M_N \!\int\!\! d^3{\bf X} 
\int\!\! dR_1 \int\!\! dR_2 \left[\phi_{T_3S_3}^{T=S}(R_2)\right]^\ast \!
\phi_{T_3S_3}^{T=S}(R_1) \!\!\!\!\!\!\!\!
\int\limits_{\;\;\;R(-T)=R_1}^{\;\;\;R(T)=R_2} \!\!\!\!\!\!\!\! D\! R \;
\mbox{Det} \left[ i\partial_t - H(U_{\rm cl}) - \Omega (t) \right]
\nonumber \\
&& \times (-i) N_c \; \mbox{Tr} \left[ R^\dagger(x^0) \Gamma R(y^0) \;
\langle {y^0,{\bf y}-{\bf X}} | \;
\frac{1}{i\partial_t - H(U_{\rm cl}) - \Omega (t)}
|{x^0,{\bf x}-{\bf X}} \; \rangle  \right] .
\;\;\;\;\;\;\;\;\;\;
\label{master}
\ee
Here $\Gamma$ denotes a matrix in Dirac spinor and isospin space, and
$\mbox{Tr}\ldots$ implies the trace over Dirac and flavor indices (the sum
over color indices has already been performed).
\par
The functional integral over rotations, $R(t)$, in 
Eq.(\ref{master}) can be computed by expanding the integrand
in the angular velocity, $\Omega$, Eq.(\ref{omega_def}), which is
of order $1/N_c$. The expansion of the determinant of the
Dirac operator gives rise to the ``kinetic term''
\be
\mbox{Det} \left[ i\partial_t - H(U_{\rm cl}) - \Omega (t) \right]
&\propto& \exp\left[ \frac{i\, I}{2}\; \int dt \; \Omega_a^2 (t) 
\; + \ldots \right] ,
\ee
where $I$ is the moment of inertia of the soliton, Eq.(\ref{I}). 
The functional integral with this action can now be computed exactly; it
corresponds to a rigid rotator described by the Hamiltonian
Eq.(\ref{H_rotator}). In addition, one has to expand the
quark Green function in Eq.(\ref{master}) in powers of $\Omega$.
The minimum power of $\Omega$ required to obtain a non-zero result 
determines the order of the matrix element in the $1/N_c$ expansion,
which is in general different for the different isospin components
of the matrix element of a given operator.
\par
{\it Isovector transversity distribution.}
We now apply the above prescription to the
calculation of the transversity distributions, 
Eqs.(\ref{delta_q_rf}, \ref{delta_qbar_def}). 
It turns out that in the case of the isovector transversity 
distribution the R.H.S.\ of Eq.(\ref{master}) is non-zero
already in zeroth order of the expansion of the quark propagator
in $\Omega (t)$, in agreement with the fact that this
isospin component is leading in the $1/N_c$--expansion. The functional 
integral over rotations in leading order of $1/N_c$ ({\it i.e.}, 
neglecting all time dependence of the collective rotation) just 
produces a delta function which enforces $R_1 = R_2$. The result
can be written in the form
\be
\delta u (x) - \delta d(x) &=&  -(2 S_3) 2 M_N
\int d^3{\bf X} \int dR \; \phi_{T_3 S_3}^\ast (R)
\; O^{I=1} ({\bf X}, R; x) \;
\phi_{T_3S_3} (R) 
\label{R_isovec}
\ee
where $O^{I=1}$ is an operator in the space of functions
of the collective coordinates
\be
O^{I=1} ({\bf X}, R; x) &=& (-i) N_c 
\int\frac{d z^0}{2\pi} e^{i x M_N z^0}
\mbox{Tr} \left[ R^\dagger \tau^3 R \;\; 
\frac{1 + \gamma^0\gamma^1}{2} \gamma_5 \gamma^3 \right.
\nonumber \\
&& \times \left. \langle z^0, {\bf z} - {\bf X} | \;
\frac{1}{i\partial_t - H(U_{\rm cl})}
| 0, -{\bf X} \; \rangle 
\right]_{z^1 = -z^0, \; z_\perp = 0} . \; 
\label{O_isovec}
\ee
The integral over rotations is readily performed; one introduces 
the rotation matrix in the adjoint representation,
\be
R^\dagger \tau^3 R &=& D_{3a}(R) \tau^a , 
\hspace{2cm} D_{ba}(R) \;\;\equiv \;\; \frac{1}{2} \mbox{Tr} 
\left( \tau_b R \tau_a R^\dagger \right) ,
\ee
and makes use of the identity
\be
\int dR \, \phi_{S_3 T_3}^\ast (R) D_{3a} (R)
\phi_{S_3 T_3} (R)  &=& -\frac{1}{3} (2 T_3 ) (2 S_3 ) \, \delta_{a3} .
\label{int_rot_isovector}
\ee
For further evaluation one writes the quark Green 
function in the static classical pion field in
Eq.(\ref{O_isovec}) in frequency representation,
\be
\langle z^0, {\bf z} - {\bf X} | \;
\frac{1}{i\partial_t - H(U_{\rm cl})}
| 0, -{\bf X} \; \rangle
   = \int \frac{d\omega}{2\pi}\; e^{-i\omega z^0}
\langle {\bf z} - {\bf X} | \;
\frac{1}{\omega  - H(U_{\rm cl})}
| -{\bf X} \; \rangle 
\label{frequency}
\ee
(the choice of contour for the $\omega$--integral will be discussed
below), and brings the matrix element into diagonal form
by introducing the finite--translation operator,
\be
\langle {\bf z} - {\bf X} | \ldots | -{\bf X} \; \rangle
&=& 
\langle -{\bf X} | \exp (i P^k z^k ) 
\ldots | -{\bf X} \; \rangle ,
\label{finite_translation}
\ee
where $P^k$ denotes the momentum operator in the space of
quark single--particle wave functions, Eq.(\ref{eigen_n}).
In this way one obtains
\be
\lefteqn{
\delta u (x) - \delta d(x) \;\; = \;\;  (2T_3 ) 
\frac{2 M_N N_c}{3} \;  (-i) \int \frac{d\omega}{2\pi} } &&
\nonumber \\
&& \times \int d^3 {\bf X} \; \mbox{Tr}\left[ \tau^3 
\frac{1 + \gamma^0\gamma^1}{2} \gamma_5 \gamma^3 
\langle -{\bf X} | \; \delta (x M_N - \omega - P^1 ) 
\frac{1}{\omega - H(U_{\rm cl})} | -{\bf X} \; \rangle \right] . \;
\label{isovector_omega}
\ee
The delta function here is the result of integrating the exponential
factors in Eqs.(\ref{O_isovec}), (\ref{frequency}) and 
(\ref{finite_translation}) over $z^0$, keeping in mind the
constraint $z^1 = -z^0$. In fact, the R.H.S.\ of Eq.(\ref{isovector_omega})
has the form of a functional trace of an operator in the space of
quark single--particle states, and can be written more concisely
as ($\mbox{Sp}$ denotes the functional trace)
\be
\delta u (x) - \delta d(x)
&=& (2 T_3 ) \frac{2 M_N N_c}{3} \; (-i) \int \frac{d\omega}{2\pi}\;
\nonumber \\
&& \times \mbox{Sp} \left[ \tau^3 
\frac{1 + \gamma^0\gamma^1}{2} \gamma_5 \gamma^3 
\delta (x M_N - \omega - P^1 ) 
\frac{1}{\omega - H(U_{\rm cl})} \right] . \;
\label{isovector_trace}
\ee
In the above expressions vacuum subtraction is implied, {\it i.e.}, one
should subtract the corresponding expressions in which the Hamiltonian
in the hedgehog pion field, Eq.(\ref{hedge}), is replaced by the
free Hamiltonian ($U = 1$).
\par
Eqs.(\ref{isovector_omega}) and (\ref{isovector_trace}) serve as the 
basis for the actual computation of the isovector distribution.
An explicit expression, suitable for numerical calculations, can be
derived by substituting in Eq.(\ref{isovector_trace}) the spectral
representation of the quark Green function in the background pion 
field,
\be
\frac{1}{\omega - H(U_{\rm cl})}
&=& 
\sum_{\scriptstyle n \atop \scriptstyle {\rm occup.}}
\frac{|n\rangle \langle n|}{\omega - E_n - i 0}
+ \sum_{\scriptstyle n \atop \scriptstyle {\rm non-occup.}}
\frac{|n\rangle \langle n|}{\omega - E_n + i 0} \; . \;
\label{spectral}
\ee
Here, $E_n$ and $|n\rangle $ are the single--particle eigenvalues and 
eigenstates of Eq.(\ref{eigen_n}). The poles are shifted to the 
upper/lower half of 
the $\omega$--plane, corresponding to whether the single--particle levels 
are occupied or not. Substituting Eq.(\ref{spectral}) into 
Eq.(\ref{isovector_trace}) we obtain a representation
of the isovector distribution as a sum over quark single
particle levels in the background pion field:
\be
\delta u (x) - \delta d(x) &=&
(2T_3 )	\frac{N_c M_N}{3} 
\left( 
\sum_{\scriptstyle n \atop \scriptstyle {\rm occup.}}
- \sum_{\scriptstyle n \atop \scriptstyle {\rm non-occup.}}
\right) 
\nonumber 
\\
&& \times
\langle n| \tau^3 \frac{1 + \gamma^0\gamma^1}{2} \gamma_5\gamma^3
\; \delta(x M_N - E_n - P^1) |n\rangle .
\label{isovector_all}
\ee
For many purposes it is desirable to have representations of the 
distribution function involving sums over only occupied, 
or only non-occupied, levels. Such representations can be
obtained from Eq.(\ref{isovector_all}) if one notes that
the sum over {\it all} levels
of the matrix element on the R.H.S.\ of Eq.(\ref{isovector_all}) 
is zero:
\be
\sum_{\scriptstyle n\;\scriptstyle {\rm all}}
\langle n| \tau^3 \frac{1 + \gamma^0\gamma^1}{2} \gamma_5\gamma^3
\; \delta(x M_N - E_n - P^1) |n\rangle  &=& 0 \;  .
\label{no_anomaly}
\ee
This condition is actually equivalent to the locality condition of the
anticommutator of quark fields in the effective low--energy theory, 
which ensures that one gets the same result 
for the distribution function in the chiral quark--soliton model if one 
starts from the QCD definition as the matrix element of the quark bilinear 
$\bar\psi (0) \ldots \psi (z)$, Eq.(\ref{delta_q_def}), or from the
equivalent definition where as the matrix element of the QCD operator where
$\bar\psi$ and $\psi$ have been anticommuted, 
$-\psi (z) \ldots \bar\psi (0)$ \cite{DPPPW96,DPPPW97}.
It is crucial that the ultraviolet regularization of the effective
low--energy theory does not destroy the property Eq.(\ref{no_anomaly}), 
as has been discussed extensively in the context of the
isoscalar unpolarized and isovector polarized distributions
in Refs.\cite{DPPPW96,DPPPW97}. A detailed investigation of the conditions 
under which Eq.(\ref{no_anomaly}) holds in the case of the isovector 
transversity distribution is presented in Subsection \ref{subsec_anomaly}
and Appendix \ref{app_anomaly}; here
we shall simply take this property for granted. In particular, 
Eq.(\ref{no_anomaly}) can be read as saying that
\be
\sum_{\scriptstyle n \atop \scriptstyle {\rm occup.}}
\langle n| \ldots |n\rangle 
&=& - \sum_{\scriptstyle n \atop \scriptstyle {\rm non-occup.}}
\langle n| \ldots |n\rangle .
\label{equivalence}
\ee
This allows to write instead of Eq.(\ref{isovector_all})
\be
\lefteqn{ \delta u (x) - \delta d(x) } && 
\nonumber \\
&=& \;\;\; (2T_3 ) \frac{2 N_c M_N}{3} \;\;\;
\sum_{\scriptstyle n \atop \scriptstyle {\rm occup.}} \;\;\;
\langle n|  \tau^3 \frac{1 + \gamma^0\gamma^1}{2} \gamma_5\gamma^3
\; \delta(x M_N - E_n - P^1) |n\rangle 
\label{isovector_occ}
\\
&=& - (2T_3 ) \frac{2 N_c M_N}{3} 
\sum_{\scriptstyle n \atop \scriptstyle {\rm non-occup.}}
\langle n|  \tau^3 \frac{1 + \gamma^0\gamma^1}{2} \gamma_5\gamma^3
\; \delta(x M_N - E_n - P^1) |n\rangle . 
\label{isovector_nonocc}
\ee
All three representations of the isovector transversity distribution,
Eqs.(\ref{isovector_all}) , (\ref{isovector_occ}) and
(\ref{isovector_nonocc}) are equivalent 
provided Eq.(\ref{no_anomaly}) holds. In all cases, the corresponding 
expressions for the antiquark distributions are obtained from those for
the quark distributions by the substitution Eq.(\ref{delta_qbar_def}).
It is understood that one should subtract from Eqs.(\ref{isovector_all}) , 
(\ref{isovector_occ}) and (\ref{isovector_nonocc}) the corresponding 
sums over eigenstates of the free Hamiltonian ($U = 1$).
\par
The expressions Eq.(\ref{isovector_occ}) and (\ref{isovector_nonocc})
can be used for numerical evaluation of the isovector distribution
function, see Section~\ref{subsec_numerical}. In particular, we shall
verify the equivalence of the two representations in the numerical 
calculations.
\par
{\it Isosinglet transversity distribution.} The calculation of the
isosinglet transversity distribution is slightly more complicated than
that of the isovector one. In the isosinglet case a non-zero result
is obtained only after expanding the integrand in Eq.(\ref{master})
to first order in the angular velocity, $\Omega$. As a consequence
this distribution is suppressed by a factor $1/N_c$ relative to the
isosinglet one, in agreement with the general $N_c$--counting arguments
of Subsection \ref{subsec_N_c}.
\par
Quark distribution functions are given by matrix elements of non-local 
operators (quark bilinears at different space--time points).
The $\Omega$--expansion of matrix elements of such operators poses
some special problems, which have been discussed extensively in
Ref.\cite{PPGWW98} in connection with the calculation of the isovector
unpolarized distribution. Terms of first order in $\Omega (t)$
arise from the first--order expansion of the Green function in 
Eq.(\ref{master}), as well as from the expansion of the structure
\[
R^\dagger(x^0) \Gamma R(y^0) ,
\] 
which is non-local in time.\footnote{If one computed not the
distribution function directly but its moments,
the second type of contribution would arise from the presence of
derivatives acting on the quark fields in the local twist--2 operators, 
which in the chiral quark--soliton model become time derivatives 
acting on the rotational matrices in 
Eq.(\ref{master}).} There are thus two types of contributions
to the distribution functions. For the isosinglet 
transversity distribution the result obtained after expanding to first 
order in $\Omega$ can be written as [{\it cf.}\ Eq.(\ref{R_isovec})]
\be
\delta u (x) + \delta d(x) &=&  -(2 S_3) 2 M_N
\int d^3{\bf X} \int dR \; \phi_{T_3 S_3}^\ast (R)
\nonumber \\
&& \times 
\; \left[ O^{I=0,\; (1)} \; + \; O^{I=0, \; (2)} 
\right]({\bf X}, R , S; x) \;
\phi_{T_3S_3} (R) ,
\label{R_isosing}
\ee
where
\be
&& O^{I=0, \; (1)} ({\bf X}, R, S; x) \;\;\; = \;\;\; (-i) N_c 
\int\frac{d z^0}{2\pi} e^{i x M_N z^0} \;
\mbox{Tr}\left[ R^\dagger R \; 
\frac{1 + \gamma^0\gamma^1}{2} \gamma_5 \gamma^3 \right.
\nonumber \\
&& \times \left. \langle z^0, {\bf z} - {\bf X} | \;
\frac{1}{i\partial_t - H(U_{\rm cl})}
\left( \frac{i}{2 I} S^a \tau^a \right)
\frac{1}{i\partial_t - H(U_{\rm cl})}
| 0, -{\bf X} \; \rangle 
\right]_{z^1 = -z^0, \; z_\perp = 0} ,
\label{O_isosing_1}
\\
&& O^{I=0, \; (2)} ({\bf X}, R, S; x) \;\;\; = \;\;\; (-i) N_c 
\int\frac{d z^0}{2\pi} e^{i x M_N z^0} \,  z^0 \;
\mbox{Tr}\left[ R^\dagger R \; \left( \frac{i}{2 I} S^a \tau^a \right)
\frac{1 + \gamma^0\gamma^1}{2} \gamma_5 \gamma^3 \right. 
\nonumber \\
&& \times \left. \langle z^0, {\bf z} - {\bf X} | \;
\frac{1}{i\partial_t - H(U_{\rm cl})}
| 0, -{\bf X} \; \rangle 
\right]_{z^1 = -z^0, \; z_\perp = 0} . 
\label{O_isosing_2}
\ee
Now the operators acting on the wave functions in collective coordinates 
involve also the
spin operator, $S$, which arises from replacing the angular velocity
according to the ``quantization rule'', Eq.(\ref{quantization_rule}).
\par
Note that in Eqs.(\ref{O_isosing_1}) and (\ref{O_isosing_2})
the spin operators do not commute with the rotational matrices, 
so in principle one should be careful about their ordering.
However, it turns out that in the case of interest here
there is no ordering ambiguity. By explicit calculation, using
the methods described in Ref.\cite{PPGWW98}, one can show that
the commutator terms corresponding to the differences between different
operator orderings give zero contribution in the final 
result.\footnote{The absence of ordering ambiguities can be shown to be 
a general feature when computing $1/N_c$--suppressed quantities, 
such as the isosinglet transversity distribution considered here, in 
{\it leading nonvanishing order}, that is, at level $\Omega^1$. 
Ordering ambiguities may arise, however, when considering $1/N_c$--corrections
to quantities which are non-zero already in leading order of the
$1/N_c$--expansion, {\it e.g.}\ in $\Omega^1$--contributions to 
the isovector transversity distribution or to the isovector longitudinally
polarized distribution. These difficulties cause a violation of the PCAC
relation and are the subject of on-going investigations.}
\par
The integrals over rotational matrices in Eqs.(\ref{O_isosing_1})
and (\ref{O_isosing_2}) are nothing but the average of the
spin operator in the rotational state,
\be
\int dR \, \phi_{S_3 T_3}^\ast (R) \; S^a \; 
\phi_{S_3 T_3} (R)  &=& S_3 \, \delta_{a3} .
\label{int_rot_isoscalar}
\ee
The further evaluation of the expressions proceeds largely in analogy 
to the isovector case. Passing to the frequency representation of
the quark Green functions, {\it cf.}\ Eq.(\ref{frequency}), and making 
use of Eq.(\ref{finite_translation}), the contributions
Eqs.(\ref{O_isosing_1}) and (\ref{O_isosing_2}) become
\be
&& \left[ \delta u (x) + \delta d(x) \right]^{(1)}
\;\; = \;\; \frac{N_c M_N}{4 I} \int \frac{d\omega}{2\pi}
\nonumber \\
&& \times \mbox{Sp} \left[ 
\frac{1 + \gamma^0\gamma^1}{2} \gamma_5 \gamma^3 
\delta (x M_N - \omega - P^1 ) 
\frac{1}{\omega - H(U_{\rm cl})} 
\tau^3 
\frac{1}{\omega - H(U_{\rm cl})} 
\right] , 
\label{isoscalar_1_trace}
\\
&&
\left[ \delta u (x) + \delta d(x) \right]^{(2)}
\;\; = \;\; \frac{N_c}{4 I} \; (-i) \frac{\partial}{\partial x} 
\int \frac{d\omega}{2\pi}
\nonumber \\
&& \times \mbox{Sp} \left[ \tau^3 
\frac{1 + \gamma^0\gamma^1}{2} \gamma_5 \gamma^3 
\delta (x M_N - \omega - P^1 ) 
\frac{1}{\omega - H(U_{\rm cl})} \right] .
\label{isoscalar_2_trace}
\ee
The derivative in $x$ in Eq.(\ref{isoscalar_2_trace}) results
from the factor of $z^0$ present in Eq.(\ref{O_isosing_2}).
One notes that this contribution is, up to a factor, equal to the 
derivative in $x$ of the leading--order ${\cal O}(\Omega^0)$ result for 
the isovector transversity distribution,
Eq.(\ref{isovector_trace}),
\be
\left[ \delta u (x) + \delta d(x) \right]^{(2)}
&=& -\frac{3}{2 I M_N} \; \frac{\partial}{\partial x}
\left[ \delta u (x) - \delta d(x) \right]^{\mbox{\scriptsize leading}} .
\label{isoscalar_2_from_isovector}
\ee
Substituting here Eqs.(\ref{isovector_occ}) or (\ref{isovector_nonocc})
one obtains a representation of this contribution as a simple sum over 
(occupied or non-occupied) quark single--particle levels.
For the contribution $(1)$, Eq.(\ref{isoscalar_1_trace}), a 
representation as a double sum over quark levels
can be derived, following the steps outlined in Appendix A of 
Ref.\cite{PPGWW98}. Again one substitutes the spectral representations,
Eq.(\ref{spectral}), for the two quark propagators. A subtle point is 
the occurrence of a double pole in the $\omega$--integral for those 
terms where the energies in the two denominators coincide, 
see Ref.\cite{PPGWW98} for details. Assuming a quasi--discrete spectrum
of levels, as appropriate for numerical calculations using a 
finite box, and separating explicitly the contributions of levels
with $E_m \neq E_n$ and $E_m = E_n$ in the double sum over levels $n$ and
$m$ one obtains a representation of the isosinglet transversity 
distribution in the form
\be
\lefteqn{ \left[ \delta u(x) + \delta d(x) \right]^{(1)} 
\;\;  = \;\; \frac{N_c M_N}{4 I} 
\left( \sum\limits_{\scriptstyle n\atop \scriptstyle{\rm occup.}}
- \sum\limits_{\scriptstyle n\atop \scriptstyle{\rm non-occup.}}
\right) } &&
\nonumber \\
&\times& \left\{
2 \sum\limits_{\scriptstyle m\atop {\scriptstyle {E_m \neq E_n}}}
\frac{1}{E_m-E_n}\; \langle n|\tau^3 |m\rangle
\langle m| \displayfrac{1+ \gamma^0 \gamma^1 }{2} 
\gamma_5 \gamma^3 \; 
\delta (E_n + P^1 - x M_N) |n \rangle 
\right.
\nonumber \\
&& - \left. 
\sum\limits_{\scriptstyle m\atop {\scriptstyle {E_m = E_n}}}
\langle n|\tau^3 |m\rangle
\langle m| \displayfrac{1+\gamma ^0\gamma^1}{2} \gamma_5 \gamma^3
\delta ^{\prime }(E_n + P^3 - x M_N)|n\rangle
\right\} .
\label{isoscalar_1_all}
\ee
As in the case of the isovector distribution, one can show that the sum 
over {\it all} levels $n$ of the terms in braces in 
Eq.(\ref{isoscalar_1_all}) is zero (see 
Subsection \ref{subsec_anomaly} and Appendix \ref{app_anomaly}),
\be
\sum_{\scriptstyle n \atop \scriptstyle {\rm all}} \;
\left\{ \ldots \right\}
&=& 0 ,
\label{no_anomaly_isoscalar}
\ee
so that we can write Eq.(\ref{isoscalar_1_all}) equivalently as
\be
\left[ \delta u(x) + \delta d(x) \right]^{(1)} 
&=& \;\;\; \frac{N_c M_N}{2 I} \;\;\;
\sum\limits_{\scriptstyle n\atop \scriptstyle{\rm occup.}}
\;\;\; \left\{ \ldots \right\}
\label{isoscalar_1_occ}
\\
&=& -\frac{N_c M_N}{2 I} 
\sum\limits_{\scriptstyle n\atop \scriptstyle{\rm non-occup.}}
\left\{ \ldots \right\} \; .
\label{isoscalar_1_non}
\ee
The expressions Eqs.(\ref{isoscalar_1_occ}) and
(\ref{isoscalar_1_non}), as well as Eq.(\ref{isoscalar_2_from_isovector})
together with Eqs.(\ref{isovector_occ}) and (\ref{isovector_nonocc}),
will be used in the numerical calculation of the isosinglet
transversity distribution in Section~\ref{subsec_numerical}.
\subsection{Ultraviolet regularization}
\label{subsec_regularization}
The effective chiral theory used to describe the nucleon
is non-renormalizable and understood to be defined with an explicit 
ultraviolet cutoff. When calculating parton distributions in this
approach one must ensure that none of their essential properties, such
as positivity, sum rules {\it etc.}, are violated
by the ultraviolet regularization. For the isoscalar unpolarized and 
isovector polarized distributions this problem has been studied in 
detail in Refs.\cite{DPPPW96,DPPPW97}. These distributions were found to 
contain ultraviolet divergences which require regularization, and it 
was shown that a possible regularization, preserving all important 
properties of the parton distributions, is by way of a Pauli--Villars 
subtraction. 
\par
We now discuss the issue of ultraviolet regularization
in the case of the transversity distributions. Our investigation consists
of two parts. First, we show that the expressions for the transversity 
distributions in the effective low--energy theory are in fact ultraviolet 
finite, and thus can in principle be computed without an ultraviolet cutoff.
Second, we investigate the role of ultraviolet regularization in the
``locality'' conditions, Eqs.(\ref{no_anomaly}) and
(\ref{no_anomaly_isoscalar}), which ensure the equivalence of the
representations of the transversity distributions as sums over occupied
and non-occupied quark levels. In Subsection \ref{subsec_anomaly}
we describe an interesting anomaly--type phenomenon found in these
sums over quark levels, which takes the form of non-commutativity
of the chiral limit and the limit of infinite ultraviolet cutoff.
\par
In order to investigate the dependence of the transversity distributions
calculated in the effective low--energy theory on the ultraviolet cutoff,
we consider the formal limit of large soliton size, in which one 
can derive explicit expressions for the distributions as functionals of the
classical pion field, $U_{\rm cl}({\bf x})$ (gradient expansion) 
\cite{DPPPW96,DPPPW97}. The gradient expansion can 
immediately be derived from the representations of the distribution
functions as functional traces with the quark Green function, 
Eq.(\ref{isovector_trace}) in the isovector, 
and Eqs.(\ref{isoscalar_1_trace}) and (\ref{isoscalar_2_trace}) in 
the isoscalar case, if one substitutes an approximate form of the
quark Green function, appropriate for small gradients of the classical
pion field, $\nabla_k U_{\rm cl}({\bf x}) \ll M$ ($k = 1, 2, 3$):
\be
&& \!\!\!\!\! \frac{1}{\omega - H(U_{\rm cl})} 
= \frac{\omega + H(U_{\rm cl})}{\omega^2 - H^2 (U_{\rm cl})}
= \frac{1}
{\omega^2 \!+\! \nabla^2 \!-\! M^2 \!-\! iM(\nablaslash U^{\gamma_5}_{\rm cl})}
\left( \omega \!-\! i\gamma^0 \gamma^k\nabla_k 
\!+\! M \gamma^0 U^{\gamma_5}_{\rm cl} \right) 
\nonumber 
\\
&& \;\; = \frac{1}{\omega^2 \!+\! \nabla^2 \!-\! M^2} \!
\sum\limits_{\scriptstyle n = 0}^\infty \!
\left[ i M (\nablaslash U^{\gamma_5}_{\rm cl})
\frac{1}{\omega^2 \!+\! \nabla^2 \!-\! M^2} \right]^n 
\; \left( \omega \!-\! i\gamma^0 \gamma^k\nabla_k 
+ M \gamma^0 U^{\gamma_5}_{\rm cl} \right) . 
\label{propagator_series}
\ee
In this formal expansion increasing numbers of gradients of 
$U_{\rm cl} ({\bf x})$
come with increasing inverse powers of $\omega$ and 
momentum, so that the leading ultraviolet divergences (if any)
of the distributions can be read off from the leading--order
gradient expansion. 
\par
{\it Isovector transversity distribution.}
In the case of the isovector transversity distribution the first
non-vanishing contribution in Eq.(\ref{isovector_trace}) comes
from the term with $n = 2$ in the expansion Eq.(\ref{propagator_series}).
Computing the resulting functional trace using a basis of plane--wave
states one obtains the leading--order gradient expansion in the
form
\be
\left[ \delta u (x) - \delta d (x) 
\right]^{\mbox{\scriptsize grad. exp.}}
&=&
\int\limits_{-\infty}^{\infty} d\xi \; e^{i\xi M_N x} \; f^{I=1} (\xi) ,
\nonumber
\ee
\be
&& f^{I=1} (\xi) = \frac{N_c M_N M}{48\,\pi^3} 
\int\limits_0^1 \!\! d\alpha 
\int\limits_0^\alpha \!\! d\beta  
\int \!\!  d^3 z \,
\mbox{tr}_{\rm fl}\left\{
U_{\rm cl}({\bf z} \!-\! \alpha \xi {\bf e}_3 ) \; 
[ \partial_i U_{\rm cl}^\dagger ({\bf z} \!-\! \beta \xi {\bf e}_3) ] \;
[\partial_j U_{\rm cl}({\bf z})] \; \tau^k \right\}
\nonumber \\
 && \;\;\;\;\;\;\;\;\;\;\;\;\;\;\;\; \times 
\left[ 
(\varepsilon^{i2j}\delta^{k1}\!+\!\varepsilon^{1ij}\delta^{k2})
+ i(\varepsilon^{1ij}\delta^{k1}\!-\!\varepsilon^{i2j}\delta^{k2}) 
\right] .
\label{isovector_gradient}
\ee
We note that in deriving this result we have assumed that the
classical pion field drops faster than $1/r^2$ for 
$r \rightarrow \infty$. This behavior is required in order
to be able to drop certain surface terms involving the
pion field at $r = \infty$, which arise from terms in 
Eq.(\ref{propagator_series}) with $n = 0$ and $1$. 
Such contributions are at the heart of the anomaly--type
phenomenon described in Subsection \ref{subsec_anomaly} 
and Appendix \ref{app_anomaly},
and we shall return to this point there.
\par
The gradient expansion, Eq.(\ref{isovector_gradient}), tells us
that the isovector transversity distribution is ultraviolet--finite:
the $\omega$--integral in Eq.(\ref{isovector_trace}) is convergent.
Thus this distribution does not require the ultraviolet cutoff of the
effective theory and can be computed in the limit of infinite cutoff.
We further note that the gradient expansion result for the 
isovector distribution is real. By explicit calculation one can show that
of $f^{I=1}(\xi)$, Eq.(\ref{isovector_gradient}) the real part is even, 
the imaginary part odd in $\xi$.
\par
{\it Isosinglet transversity distribution.} 
In a similar way one can 
derive the gradient expansion for the isosinglet transversity 
distribution. (We do not quote the expressions here.) The gradient
expansion shows that this distribution is finite in the limit of large
UV cutoff and does not require regularization.
\subsection{Anomaly--type phenomenon in the tensor charge}
\label{subsec_anomaly}
The tensor charges, and, more generally, the transversity
distributions show an interesting anomaly--type phenomenon, which
we shall discuss now. Aside of being of general interest, this
anomaly has direct implications for our procedure of calculating
the transversity distributions in the chiral quark--soliton model,
as it is related to the equivalence of the representations of the
distributions as sums over occupied or non-occupied quark levels,
Eqs.(\ref{isovector_occ}) and (\ref{isovector_nonocc}). This is
equivalent to the condition that the sum of the single--particle
matrix elements over all levels ({\it i.e.}, occupied and
non-occupied) be zero, Eq.(\ref{no_anomaly}). A thorough
understanding of this phenomenon is thus necessary for a reliable
calculation of the distributions. In this section we illustrate
the essential points by considering the simplest case of the
isovector tensor charge; the corresponding calculation for the
$x$--dependent distribution function presents only technical
difficulties and is presented in Appendix \ref{app_anomaly}.
\par
{\it Anomaly in the tensor charge.} At its most general, the
anomaly we are dealing with is a statement about the functional
trace of a chirally odd operator in the space of quark
single--particle wave functions in a chiral background field. Let
us consider the following functional trace: \beq \mbox{Sp} \left[
\tau^a \gamma_0\ i \sigma_{0k}\ \gamma_5 \right]\, ,
\label{spur:naive} \eeq which appears in the expression for the
tensor charge. Naively this trace would be zero because the
corresponding matrix traces in flavor and spin spaces are zero.
However, actually we deal here with an uncertainty of the type
$0\cdot\infty$. To resolve this uncertainty one has to introduce a
regularization. We choose here for illustrative purposes to
regularize the trace (\ref{spur:naive}) in the following way:

\beq \mbox{Sp} \left[ \tau^a \gamma_0\ i\sigma_{0k}\ \gamma_5
\right]_{\rm reg} =\lim_{\epsilon\to 0}\mbox{Sp} \left[ \tau^a
\gamma_0\ i\sigma_{0k}\ \gamma_5 e^{-\epsilon H^2}\right] \, .
\label{spur:regul} \eeq Here \beq H^2(U) \;\; = \;\;
 -\nabla^2 + M^2 + i M (\nablaslash U^{\gamma_5}) \, ,
\eeq is the Hamiltonian squared of the chiral quark-soliton model.
In the limit of small $\epsilon$ the expression in the RHS of
eq.~(\ref{spur:regul}) can be computed analytically with the help
of semiclassical expansion. The leading contribution has the
following form:

\beq \lim_{\epsilon\to 0}\mbox{Sp} \left[ \tau^a \gamma_0\
i\sigma_{0k}\ \gamma_5 e^{-\epsilon H^2}\right]= \frac{i M}{2\pi
\sqrt{\pi \epsilon}} \int d^3 x\ \partial_k\ \mbox{tr}\left[\tau^a
U({\bf x}) \right]\, . \eeq We see that the trace (\ref{spur:regul}) is
linearly divergent provided the chiral field $U({\bf x})$ falls
off not too rapidly at spatial infinity. For the soliton solution
in the chiral limit the chiral field behaves at the spatial
infinity as $U({\bf x})-1\sim \frac{x^a\tau^a}{|{\bf x}|^3}$ what
leads to a non-zero coefficient in front of the linear divergence
$1/\sqrt{\epsilon}$. The corresponding coefficient can be computed
in terms of axial charge of the nucleon $g_A$ keeping in mind the
asymptotics of the soliton solution in the chiral limit: \beq
U({\bf x})\sim 1-i\frac{3 g_A }{8\pi F_\pi^2}\
\frac{x^a\tau^a}{|{\bf x}|^3}\,\label{g_A_from_kappa}.\eeq
The result has the form:
\beq \lim_{\epsilon\to 0}\mbox{Sp} \left[ \tau^a \gamma_0\
i\sigma_{0k}\ \gamma_5 e^{-\epsilon H^2}\right]= \frac{g_A
M}{2\pi F_\pi^2 \sqrt{\pi \epsilon}}\ \delta^{a}_{k}\, .
\label{A_linear_divergence} \eeq Note however that in the case of
non-zero quark masses (whatever small they are) the linear
divergence and hence the anomaly is zero. For the calculations in
the finite volume the condition for the absence of the anomaly is
$M_\pi\times$box size$\gg 1$. When computing the tensor charge (or
the transversity distribution) by summing over quark levels in a
finite volume, in the strict chiral limit one would find that the
summation over occupied levels, Eq.(\ref{isovector_occ}), and
non-occupied levels, Eq.(\ref{isovector_nonocc}), does not give
equivalent results; rather, the difference diverges linearly with
the ultraviolet cutoff. This is indeed what we observe in the
numerical calculations.
\par
{\it Elimination of the anomaly in the numerical calculations.} In
the numerical calculations, which are based on diagonalization of
the Dirac Hamiltonian in a finite box, it is important to
eliminate the ``anomaly'', Eq.(\ref{A_linear_divergence}). A way
to do this is to modify the soliton profile at large distances in
a way that it vanishes faster than $1/r^2$. For example, one may
use a modified version of the variational profile suggested in
Ref.\cite{DPP88},
\be
P_{M_\pi} (r) &=& -2\;\arctan\left[ \frac{r_0^2}{r^2} (1 + M_\pi r
) \exp (-M_\pi r) \right] , \label{var_mpi} \ee where $r_0 \approx
1.0 M^{-1}$ is the usual soliton size parameter. For $M_\pi \neq
0$ this profile decays exponentially at large $r$, while for small
$r$ it differs only very little from the massless profile. The
``anomaly'', Eq.(\ref{A_linear_divergence}), can then be made
arbitrarily small by choosing the radius of the box used in the
numerical calculation, $D$, sufficiently large so that \beq M_\pi
D \gg 1. \eeq The limit $M_\pi \rightarrow 0$ is then taken at the
very end of the calculation, by extrapolation of the numerical
data. A measure of the successful elimination of the ``anomaly''
in the numerical calculations is the equivalence of the results
obtained summing over occupied and non-occupied states. In the
actual calculations we have used values of $M_\pi$ of the order of
$0.3 \ldots 0.6\, M$, and a box radius of $20\, M^{-1}$. For these
parameters we observed equivalence of summing over occupied and
non-occupied states at the level of $1\ldots 2 \%$.
\section{Results}
\setcounter{equation}{0}
\label{sec_results}
\subsection{Transversity vs.\ longitudinally polarized distributions}
\label{subsec_transverse_vs_longitudinal}
Having derived the expressions for the transversity distributions
in the effective low--energy theory, we now proceed to compute the
distributions and discuss their properties. Before turning to the
numerical evaluation of the expressions it is instructive to compare
the expressions for the transversity quark-- and antiquark distributions 
with those of the corresponding 
longitudinally polarized distributions 
at a qualitative
level. In Section~\ref{subsec_numerical} we shall then compare
the numerical results.
\par
Since the nucleon's axial and tensor charges are the same in the
non-relativistic quark model, it is generally thought that 
differences between the two distributions are a measure of
``relativistic effects'' in the nucleon.
The chiral--quark soliton model, which is a fully relativistic 
description of the nucleon, offers the unique possibility to 
study ``relativistic effects'' in the quark-- and antiquark 
distributions in a controlled way. The relevant parameter here
is the ``soliton size'', {\it i.e.}, the radius of the classical
pion field of the nucleon. Although in reality the soliton 
size is determined by the minimization of the classical energy, 
Eq.(\ref{E_tot}), it is instructive to regard it as a parameter 
and to study the dependence of quantities such as the quark and 
antiquark distribution functions on it \cite{Blotz95}. For small 
soliton sizes the bound--state level of quarks becomes weakly 
bound, and the lower Dirac components in its wave function 
become small. It was shown that in this limit nucleon matrix elements 
of a variety of local operators tend to their corresponding values 
in the non-relativistic quark model \cite{Blotz95}. On the other
hand, in the limit of large soliton sizes the bound state level 
approaches the negative continuum, and one may perform an expansion of nucleon 
matrix elements in inverse soliton size, which technically is obtained 
by expanding in gradients of the classical pion field (gradient 
expansion). In this limit this picture of the nucleon shows many 
similarities with the Skyrme model. In this sense one may say that
the chiral quark--soliton model interpolates between the non-relativistic
quark model and a Skyrme soliton picture of the nucleon.
\par
{\it The quark model limit (small soliton size).}
It is thus interesting to study the transversity and longitudinally 
polarized distributions in dependence on the soliton size. Consider
first the isovector distributions, which are leading in the $1/N_c$--expansion
and both given by simple sums over quark single--particle levels, 
see Eq.(\ref{isovector_occ}) and Ref.\cite{DPPPW96}. For small
soliton size the contributions due to the polarized Dirac continuum become 
negligible, and both sums are dominated by the contribution from of
the bound--state level:
\be
\lefteqn{ 
\left. \begin{array}{r} 
[\delta u (x) - \delta d(x)]_{\rm lev} \\[2ex]
[\Delta u (x) - \Delta d(x)]_{\rm lev}
\end{array} \right\} } && 
\\ 
&=&  (2T_3 ) \, \frac{2 N_c M_N}{3} 
\int\!\!\frac{d^3k}{(2\pi)^3} \;
\Phi_{\rm lev}^\dagger ({\bf k})
\left\{ \begin{array}{r} 
\tau^3 \displayfrac{1 + \gamma^0\gamma^1}{2} \gamma_5\gamma^3 
\delta(x M_N - E_n - k^1 )
\\[2ex]
(-) \tau^3 \displayfrac{1 + \gamma^0\gamma^3}{2} \gamma_5 
\delta(x M_N - E_n - k^3 )
\end{array} \right\}
\Phi_{\rm lev} ({\bf k}) .
\nonumber 
\ee
Evaluating this contribution using the explicit form of the
bound--state level wave function \cite{DPP88} one finds
\cite{DPPPW96,PP96}:
\be
\left[ \, \delta u(x) \, - \, \delta d(x) \, \right]_{\rm lev}
&=& \frac{N_c\Mn}{3} \!\! 
\int\limits_{|x\Mn-E_{\rm lev}|}^\infty\!\frac{{\rm d}k}{2k}\;
\Biggl\{h(k)-j(k)\,\frac{x\Mn-E_{\rm lev}}{k}\Biggr\}^2
\\ 
\left[ \, \Delta u(x) \, - \, \Delta d(x) \, \right]_{\rm lev} 
&=& \frac{N_c M_N}{3}
\!\!\int\limits_{|xM_N - E_{\rm lev}|}^{\infty}\!\!
\frac{{\rm d}k}{2k}\;
\Biggl\{ h^2(k) +\left[2 \frac{(xM_N - E_{\rm lev})^2}{k^2}
- 1 \right] j^2(k)
\nonumber \\
&&
- 2 \frac{(xM_N - E_{\rm lev})}{k} h(k) j(k)\Biggr\},
\ee
where $h (k)$ and $j(k)$ are the Fourier transforms of the radial wave 
functions corresponding to the upper and lower components of the
Dirac spinor wave function (see Appendix B of Ref.\cite{DPPPW96} for 
details). From the Dirac equation in the hedgehog pion field it 
follows that in the limit of small soliton size the lower Dirac component
of the level wave function becomes small: $j(k) \rightarrow 0$, so that
the bound state effectively becomes non-relativistic. One sees
that in this case the expressions for the isovector transversity and 
longitudinally polarized distributions coincide. 
\par
In a similar way one may investigate the limit of small soliton size in 
the isoscalar polarized distributions. This is slightly more complicated,
as these distributions are given by double sums over quark levels.  
Consider the representation of the isoscalar distribution in the form
of a spectral integral, Eqs.(\ref{isoscalar_1_trace}) and
Eq.(\ref{isoscalar_2_trace}). 
One can show that in the limit of small soliton size
the dominant contribution is the sum of two contributions which one 
obtains by replacing one of the propagators in Eqs.(\ref{isoscalar_1_trace}) 
by the pole associated with the bound--state level, $|\mbox{lev} \rangle 
\langle \mbox{lev}| / (\omega - E_{\rm lev})$, the other one by the free
propagator, $1/(\omega - M)$. 
In this way one can show that also in the isoscalar case the transversity and 
longitudinally polarized contributions tend to the same function in the 
``quark model'' limit.
\par
{\it The ``skyrmion'' limit (large soliton size).} 
In the formal limit
of large soliton size the bound--state level of quarks disappears in the 
negative--energy Dirac continuum. In this limit functions of the soliton
field can be computed by expanding in gradients of the pion field; 
more precisely in the parameter $\partial U / M$. The techniques for
deriving the gradient expansion of quark and antiquark distributions 
have been developed in Refs.\cite{DPPPW96,DPPPW97}. For the isovector 
transversity distribution the result of the leading--order gradient 
expansion has been given in Eq.(\ref{isovector_gradient}).
The corresponding expression for the isovector longitudinally polarized 
distribution has been derived in 
Ref.\cite{DPPPW96}\footnote{The isovector longitudinally polarized 
quark-- and antiquark distributions coincide only in the hypothetical 
limit of large soliton size. For finite soliton sizes they are, of course,
different.}:
\be 
\lefteqn{
\left. \begin{array}{c}
\Delta \bar u(x) - \Delta \bar d(x) \\
\Delta u(x) - \Delta d(x)
\end{array} \right\} }
&& \nonumber \\
&\sim&
\frac{F_\pi^2 M_N}{3} \int_{-\infty}^\infty \frac{d\xi}{2\pi}
\frac{\cos M_N \xi x}{\xi} \;
\int d^3 y \; {\rm tr}\,\left[ \tau^3 (-i) 
U_{\rm cl} ({\bf y} + \xi {\bf e}_3)
U_{\rm cl}^\dagger ({\bf y}) \right] 
\label{isovector_longitudinal_gradient} ,
\ee
where $F_\pi$ is the pion decay constant,
\be
F_\pi^2 = 4N_c\!\int\!\!\frac{d^4k}{(2\pi)^4}\;\frac{M^2}{(M^2 + k^2)^2}-
\frac{M^2}{M_{PV}^2\!\!\!}\,\,
4N_c\!\int\!\!\frac{d^4k}{(2\pi)^4}\;\frac{M^2_{PV}}{(M^2_{PV} + k^2)^2}
= \frac{N_c M^2\!\!}{4\pi^2} \log \frac{M_{PV}^2}{M^2} .
\label{fpi}
\ee
(In the last line we have given the result in the case when the 
logarithmic divergence of the integral is regularized by a 
Pauli--Villars subtraction.) Comparing the two expressions we see that in 
the limit of $r_0 \rightarrow \infty$ ($r_0$ is the parameter characterizing 
the soliton size) the isovector transversity distribution, 
Eq.(\ref{isovector_gradient}), is suppressed relative to the 
longitudinally polarized one, Eq.(\ref{isovector_longitudinal_gradient}),
by a factor of $1/(M r_0)$. Thus we see that in this limit the two 
distributions behave differently at a qualitative level. This is not 
unexpected, since, as explained above, the limit of large soliton size
can be regarded as the opposite of the non-relativistic limit.
\par
In the previous calculation of the isovector polarized quark distributions 
in the chiral quark--soliton model \cite{DPPPW96,DPPPW97} it was observed 
that the gradient expansion expression, when evaluated at the the 
physical soliton size, gives a fairly realistic description of the 
isovector polarized {\it antiquark} distribution. (The {\it quark} 
distribution, in contrast, is poorly described by the gradient expansion
expression.) The same can be expected to apply to the isovector 
transversity distribution. If this is so, it implies that we should 
expect the isovector transversity antiquark distribution to be smaller
than the corresponding longitudinally polarized one, since the latter
is parametrically suppressed in the soliton size. This expectation 
is indeed borne out by the numerical results, as will be shown in 
Section~\ref{subsec_numerical}.
\par
The behavior of the isosinglet polarized distribution in the 
limit of large soliton size can also be studied using gradient 
expansion. The gradient expansion is readily derived from the
spectral representation of the distribution functions, 
Eqs.(\ref{isoscalar_1_trace}) and Eq.(\ref{isoscalar_2_trace}),
by formally expanding the quark propagators in powers of derivatives 
of the pion field, as discussed in Section~\ref{subsec_regularization},
{\it cf.}\ Eq.(\ref{propagator_series}). We shall not quote the 
lengthy expressions here. Rather, we shall discuss the properties
of these distributions at the hands of the exact numerical results
in Section~\ref{subsec_numerical}.
\subsection{Numerical results}
\label{subsec_numerical}
We now turn to the numerical evaluation of the transversity distributions
in the chiral quark--soliton model. The numerical calculations are
based on the expressions of the transversity quark-- and antiquark 
distributions as sums over quark levels in the classical pion field of
the soliton, Eq.(\ref{isovector_all}) for the isovector, 
and Eq.(\ref{isoscalar_1_all}) for the isosinglet distribution. In fact,
in the actual calculations we prefer to use the representations of
the distributions as sums over either occupied or non-occupied levels,
Eqs.(\ref{isovector_occ}) and (\ref{isovector_nonocc})
{\it viz.}\ Eqs.(\ref{isoscalar_1_occ}) and (\ref{isoscalar_1_non}); 
the check of 
equivalence of summation over occupied or non-occupied states offers
a very powerful check of the numerical procedure.
\par
{\it Numerical method.}
The numerical method for evaluating the sums over single--particle
quark levels has been described in Ref.\cite{DPPPW97}; see also 
Ref.\cite{PPGWW98}. The spectrum of quark levels is made quasi--discrete
by placing the soliton in a spherical 3--dimensional box, imposing
so-called Kahana--Ripka boundary conditions \cite{KR84}. The eigenvalues
and eigenfunctions of the Dirac Hamiltonian in the background pion field 
are then found by numerical diagonalization, using the eigenstates of the
free Hamiltonian (zero pion field) as a basis. One can then compute the
sum over quark levels, using the fact that the matrix elements of the 
single--particle operators appearing in Eqs.(\ref{isovector_all})
and (\ref{isoscalar_1_all}) between the basis states are either
known or can easily be computed numerically (see Ref.\cite{DPPPW97} for 
details). 
\par
Since the Hamiltonian is invariant under combined spin--isospin 
rotations, and the basis states are eigenstates of the sum of the total 
angular momentum and isospin, it is advantageous to convert the 
expressions for the distribution functions, Eqs.(\ref{isovector_all})
and (\ref{isoscalar_1_all}), to a spherically symmetric form before
performing the sums over levels. In this way one greatly reduces the 
number of non-zero matrix elements of the relevant single--particle 
operators between basis states. This symmetrization  is achieved
by replacing in Eqs.(\ref{isovector_all}) and (\ref{isoscalar_1_all})
the $1$-- and $3$-- vector components of the momentum operator, 
gamma matrices, and isospin matrices, by components along two 
orthogonal 3--dimensional unit vectors, ${\bf n}$ and ${\bf m}$, and 
averaging over their orientations, with the constraint
${\bf n}\cdot{\bf m} = 0$. The matrix elements of the resulting 
``spherically symmetric'' single--particle operators between basis
states of angular momentum plus isospin can then 
easily be computed.
\par
In the calculation of quark-- and antiquark distributions in the chiral
quark--soliton model an additional complication arises due to the fact 
that the relevant single--particle operators are discontinuous
functions of the single--particle momentum and energy operators,
which leads to problems when performing the sums over of quark levels
in a discrete basis. It was shown in Ref.\cite{DPPPW97} that this
problem can easily be circumvented by applying Gaussian smearing
in the variable $x$ to expressions for the distribution functions.
We shall apply this method here, using a smearing width of 
$\gamma = 0.1$, see Ref.\cite{DPPPW97} for details.
\par
In the numerical calculations we have used the soliton profile determined
in a self--consistent minimization of the soliton energy calculated with
a Pauli--Villars cutoff \cite{WG97}. In the case of the isosinglet
unpolarized distribution this choice of profile, combined with 
a corresponding Pauli--Villars regularization of the distribution
functions, allowed to preserve the momentum sum rule for the flavor--singlet
quark plus antiquark distributions. In the case of
the transversity distributions, as shown in 
Subsection \ref{subsec_regularization}, the expressions for the 
distribution functions are ultraviolet finite, and there are no
sum rules linking them to any quantity requiring regularization. 
For reasons of consistency in the choice of
model parameters we compute also these distributions with the
soliton profile, nucleon mass, and moment of inertia, calculated
with the Pauli--Villars regularization of Ref.\cite{WG97}.
\par
Although the transversity distributions are ultraviolet finite, 
numerical calculations require that we first evaluate the sums over 
levels,  Eqs.(\ref{isovector_all}) and (\ref{isoscalar_1_all}), applying
some smooth cutoff for levels with large energies, which at the end of 
the calculation is removed by extrapolation to infinity.
When performing the numerical calculations in a finite--size box, it is 
important to take into account the ``anomaly--type'' phenomenon, described 
in detail in Appendix \ref{subsec_anomaly}. The nature of this phenomenon 
is non-commutativity between the limit of infinite energy cutoff
and the chiral limit. In the strict chiral limit, where the soliton
profile falls off like $P(r) \sim 1/r^2$ for $r \rightarrow \infty$, 
the sums over occupied and non-occupied quark levels would no longer
be equivalent. For this reason it is essential to perform the numerical
calculations in the box with a soliton profile of finite range; for
example, with a profile falling off as $\exp (- M_\pi r)/r$ for
$r \rightarrow \infty$, where $M_\pi$ is a parameter (not necessarily
equal to the physical pion mass), and extrapolate to $M_\pi \rightarrow 0$
at the end. The condition which must be satisfied for a proper calculation
in a finite box is $M_\pi D \gg 1$, where $D$ is the box radius
(see Appendix \ref{subsec_anomaly}). If this condition is satisfied
one observes equivalence of the results of summation over occupied
and non-occupied states in the numerical calculations. In the actual 
calculations we have used a box size of 
$D = (20 \ldots 30) M^{-1}$ and values of $M_\pi$ in the 
range $(0.3 \ldots 0.6) M$.
\par
\begin{figure}[t]
\begin{center}
\includegraphics[width=9.5cm,height=9.5cm]{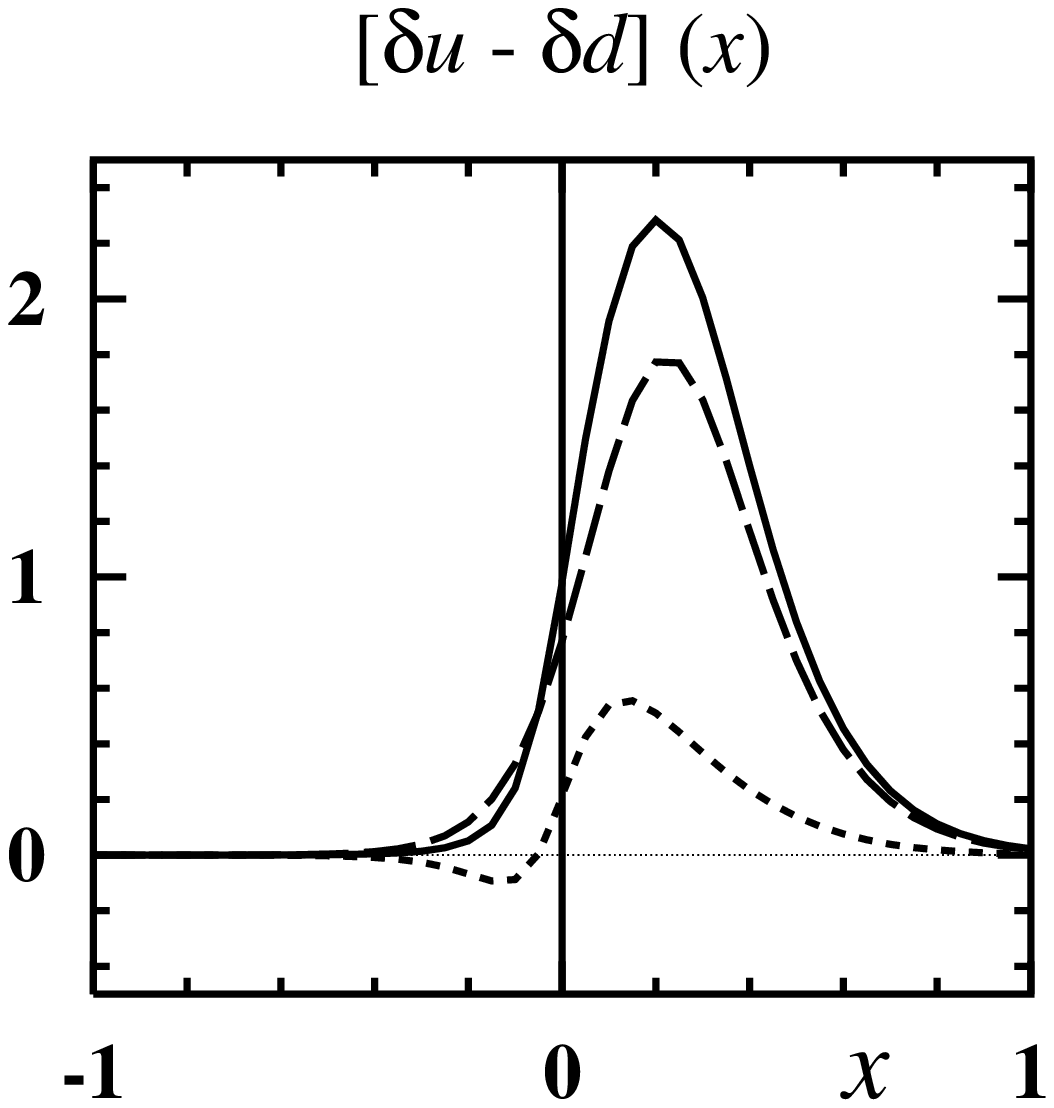}
\\
\includegraphics[width=9.5cm,height=9.5cm]{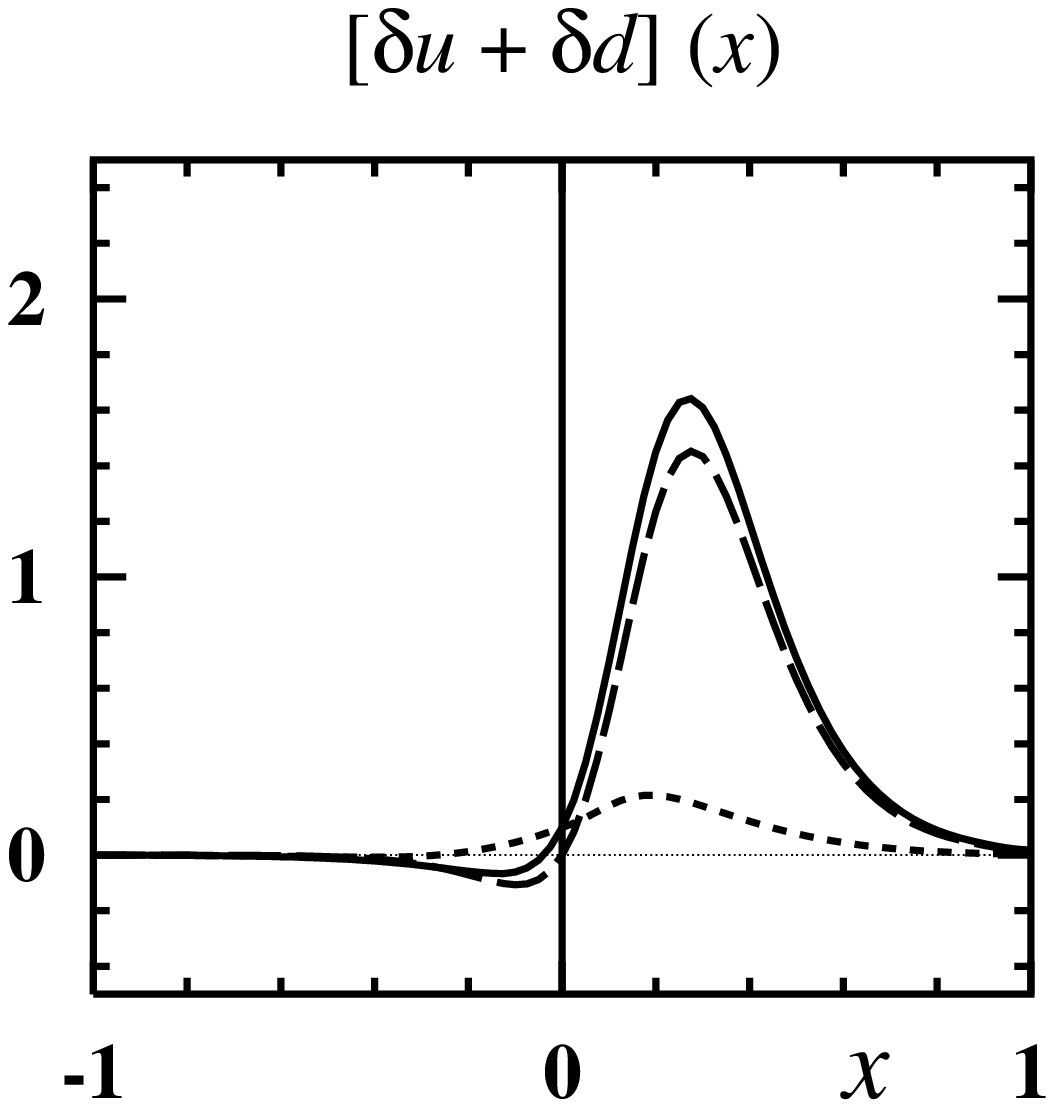}
\end{center}
\caption[]{\it The isovector (top) and isoscalar (bottom) transversity 
quark-- and antiquark distributions obtained from the chiral 
quark--soliton model. The functions shown here represent
$[\delta u \mp \delta d](x)$ at $x > 0$ and 
$-[\delta \bar u \mp \delta \bar d] (-x)$ at $x < 0$.
\underline{Dashed lines:} Contributions of the bound--state level.
\underline{Dotted lines:} Contributions of the Dirac continuum.
\underline{Solid lines:} Total results (sums of bound--state level
and Dirac continuum).}
\label{fig_delta}
\end{figure}
\begin{figure}[t]
\begin{center}
\begin{tabular}{cc}
\includegraphics[width=7.4cm,height=7.4cm]{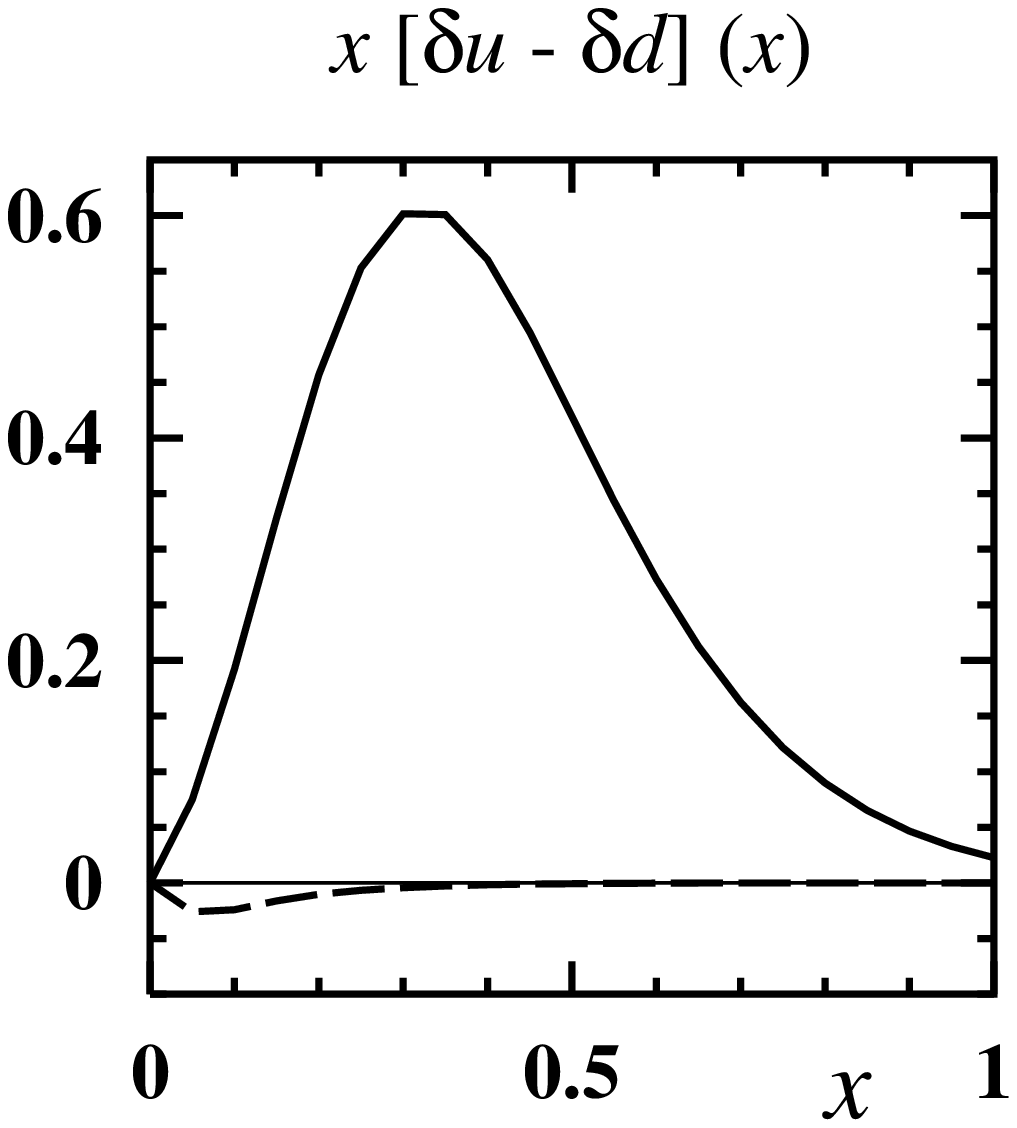}
&
\includegraphics[width=7.4cm,height=7.4cm]{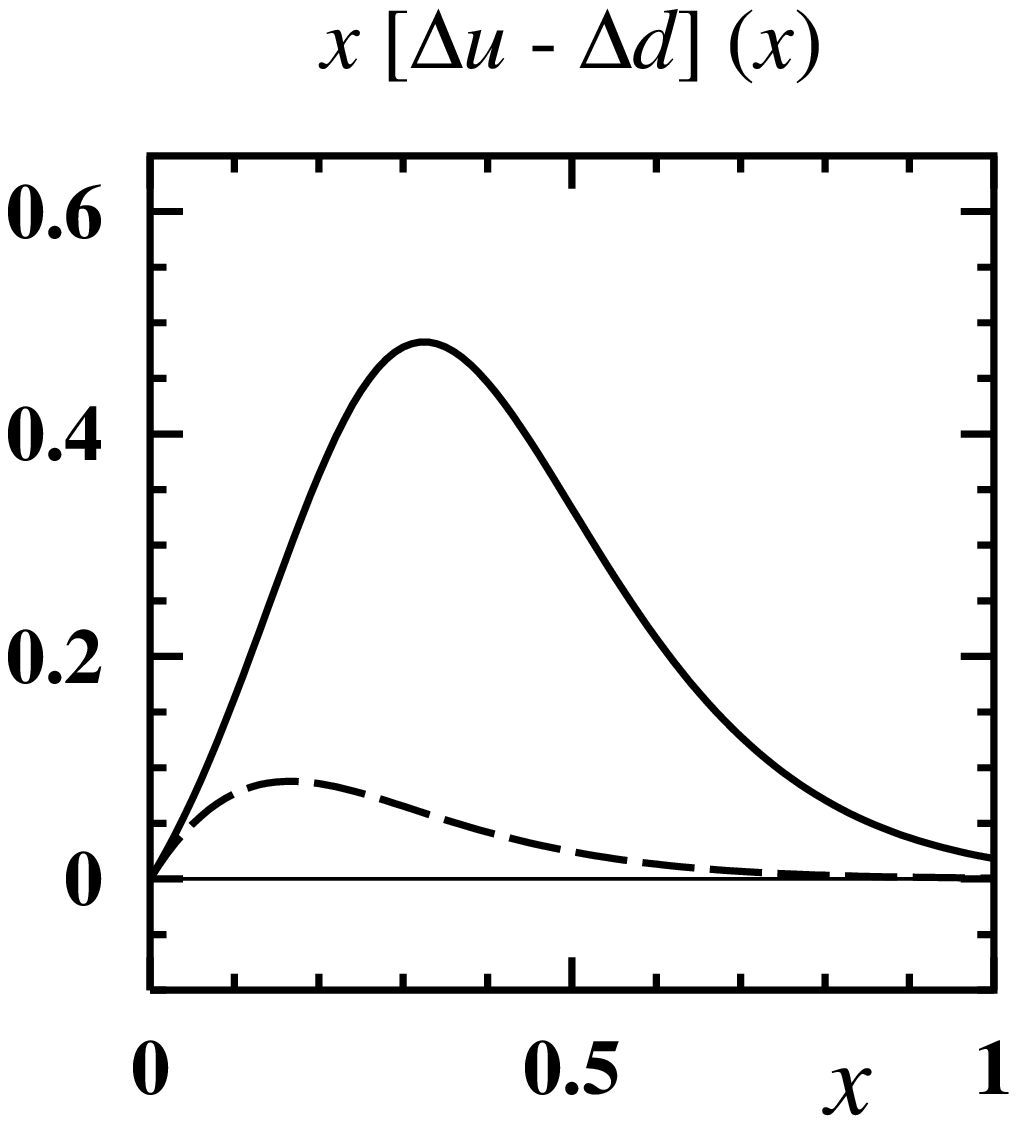}
\\
\includegraphics[width=7.4cm,height=7.4cm]{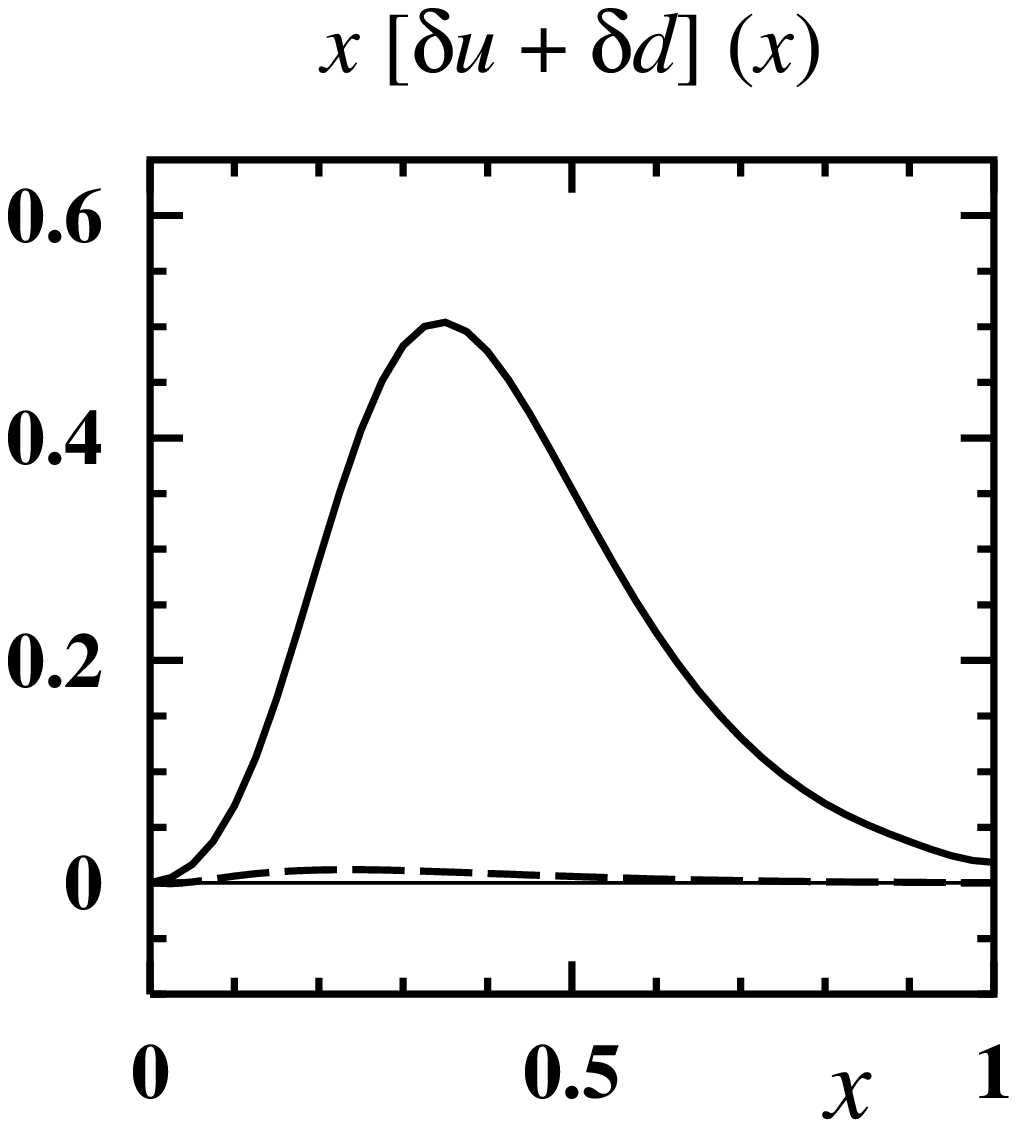}
&
\includegraphics[width=7.4cm,height=7.4cm]{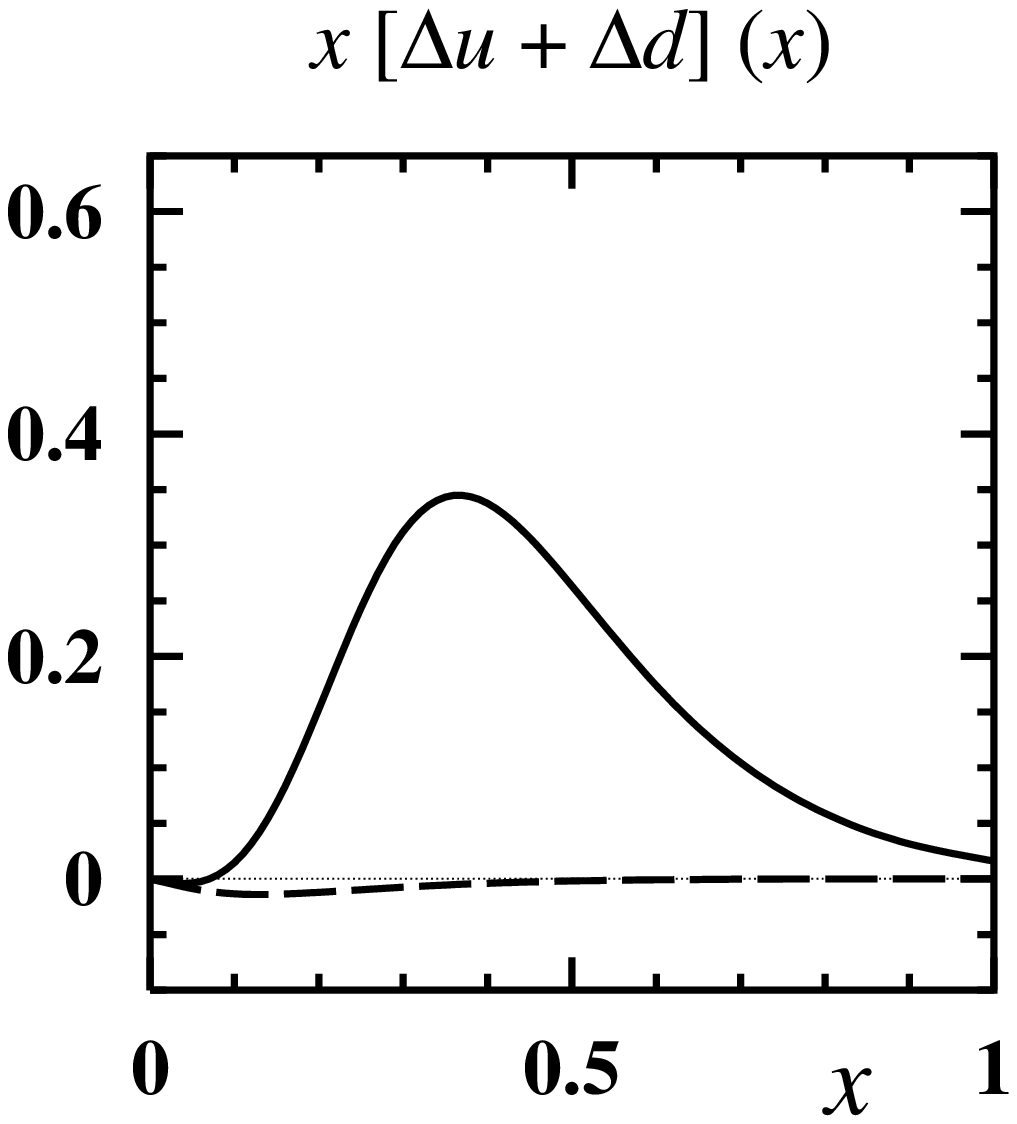}
\end{tabular}
\end{center}
\caption[]{\it The total isovector (top row) and isoscalar 
(bottom row) transversity and longitudinally polarized 
quark-- and antiquark distributions, multiplied by $x$. 
Shown are the total results (sum of level and continuum 
contributions), corresponding to the solid lines in 
Fig.~\ref{fig_delta}.) \underline{Solid lines:} Quark distributions.
\underline{Dashed lines:} Antiquark distributions.}
\label{fig_xdelta}
\end{figure}
\begin{figure}[t]
\begin{center}
\begin{tabular}{cc}
\includegraphics[width=7.4cm,height=7.4cm]{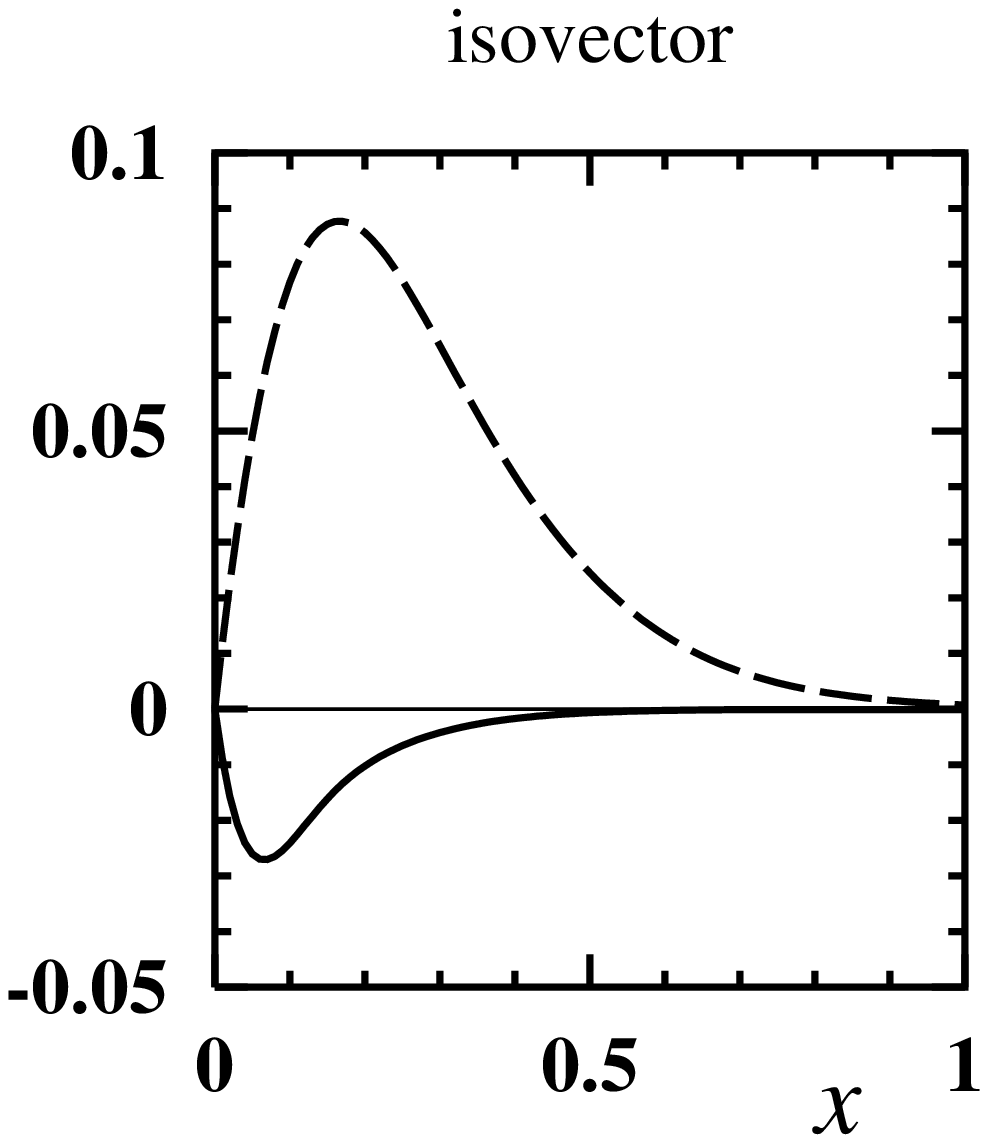}
&
\includegraphics[width=7.4cm,height=7.4cm]{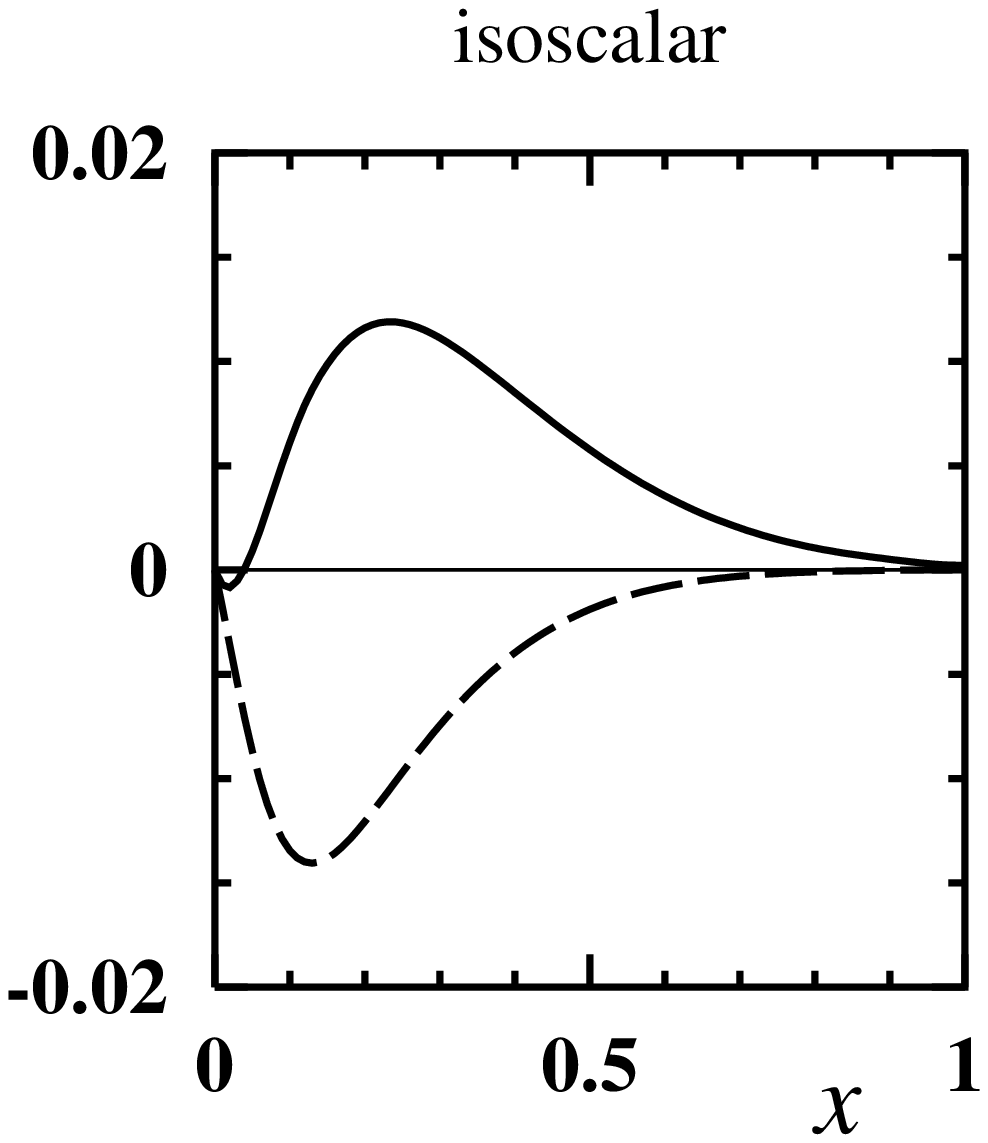}
\end{tabular}
\end{center}
\caption[]{\it The transversity and longitudinally polarized 
antiquark distributions, see Fig.~\ref{fig_xdelta}. 
Left: Isovector distributions, 
$x[\delta \bar u - \delta \bar d](x)$
and $x[\Delta \bar u - \Delta \bar d](x)$.
Right: Isoscalar distributions, 
$x[\delta \bar u + \delta \bar d](x)$
and $x[\Delta \bar u + \Delta \bar d](x)$.
\underline{Solid lines:} Transversity antiquark distributions
\underline{Dashed lines:} longitudinally polarized antiquark 
distributions.}
\label{fig_xanti}
\vspace{10cm}
\end{figure}

{\it Numerical results.}
The numerical results for the isovector transversity distributions are
shown in the upper plot of 
Fig.\ref{fig_delta}. It is interesting to first compare the
relative contributions of the bound-state level (dashed line), and of the 
Dirac continuum (dotted line) of quarks, to the quark-- and antiquark 
distributions. Fig.\ref{fig_delta} shows that the bound--state level 
gives the numerically dominant contribution 
to both the quark and antiquark transversity distributions. This is in
agreement with the observation that the isovector distribution is
suppressed in gradient expansion, since the gradient expansion expression
may be regarded as an estimate of the Dirac continuum contribution.
Our results justify the approximation made by some of us in Ref.\cite{PP96}, 
where only the level contribution to the isovector transversity distributions
was retained. Fig.\ref{fig_xdelta} (right column) shows the isovector 
quark-- and antiquark distributions multiplied by $x$. The antiquark 
distributions are shown separately in Fig.~\ref{fig_xanti}. Note that our 
approach predicts a definite sign for the isovector antiquark distribution,
$\delta \bar u (x) - \delta \bar d (x) < 0$. In the numerical calculation
at the physical soliton radius this happens because of the dominance
of the level contribution, which has positive sign, see 
Fig.\ref{fig_delta}. The gradient expansion, Eq.(\ref{isovector_gradient}),
which becomes exact in the limit of large soliton size, also predicts
a negative sign of $\delta \bar u (x) - \delta \bar d (x)$. 
As it is insensitive to the soliton size, the sign of the isovector
antiquark distribution can be taken as a robust prediction of this model.

\par
The different contributions to the isoscalar transversity distribution 
are shown in the lower plot of Fig.\ref{fig_delta}. 
One sees that in the isoscalar case the 
dominance of the level contribution is even more pronounced than in the 
isovector case. In particular, in the isoscalar antiquark distribution
the level and continuum contributions all but cancel, leaving the
total isoscalar antiquark distribution to be much smaller than the 
quark distribution (by a factor of the order of 1--2$\times 10^{-2}$, 
the precise value depending on $x$).
\par
{\it Comparing with the longitudinally polarized distributions.}
It is interesting to compare the numerical results for the transversity
distributions with those of the longitudinally polarized 
distributions \cite{Goeke:2000ym,Goeke:2000wv}, which are shown in 
the right column of Fig.~\ref{fig_xdelta}. One notices that, generally 
speaking, the quark distributions 
are of similar shape, although of different magnitude. The ratios
of the numerical results for the transversity to the longitudinally polarized 
quark distributions are well described by the forms
\be
\frac{\delta u(x) - \delta d(x)}{\Delta u(x) - \Delta d(x)}
&\approx & 1.25,
\label{ratio_isovector}
\\
\frac{\delta u(x) + \delta d(x)}{\Delta u(x) + \Delta d(x)}
&\approx & 2.0 - 1.5 \, x \; .
\label{ratio_isoscalar}
\ee
These fits apply for values of $x > 0.1$. (Note that our model,
at the present level of approximation, cannot be applied to study the 
small--$x$ behavior of the parton distributions at the low scale. 
Anyway, it is the intermediate-- and large--$x$ region of the distributions at
the low scale of the model which determines the behavior of the distributions 
in the small--$x$ region at experimentally relevant scales due to perturbative 
evolution.)
Note that this parameterization is a purely numerical fit. We find it 
useful to represent the results in this way, since we believe the ratios 
of distributions to be less model--dependent than the distributions
themselves. In our estimates of spin asymmetries in 
Section~\ref{sec_asymmetries}
we shall use these ratios, together with a parameterization of the
longitudinally polarized distributions obtained from fits to inclusive
DIS data.
\par
Comparing the transversity and longitudinally polarized antiquark distributions
we see that they are very different. This is in accordance with the fact
that the two isovector distributions appear in different orders of the 
gradient expansion, see Section~\ref{subsec_regularization}. 
It would thus not be useful
to parameterize their ratio, the more since the standard parameterizations
for the longitudinally polarized distributions 
exhibit large differences in the polarized 
antiquark distributions\footnote{The flavor asymmetry 
of the polarized antiquark distributions, $\Delta \bar u - \Delta \bar d$,
was assumed to be zero in most parameterizations, e.g. \cite{GRSV96}.}. 
When computing
observables one should rather use the results for these distributions
directly.
\par
{\it Results for tensor charges}. The tensor charges can be obtained
by integrating the numerical results for the transversity distributions,
or directly by computing the matrix elements of the local 
chirally--odd operators, Eq.(\ref{tensor_def}); both calculations
give identical results. Our numerical results for the isovector
and isoscalar tensor charges are
\be
\delta u - \delta d &=& 1.06 , \\
\delta u + \delta d &=& 0.63 .
\ee
These numbers refer to the low normalization point of 
${\cal O}(600 \, {\rm MeV})$; however, the scale dependence of these
quantities is known to be weak. Our results are consistent
with those of previous calculations in the chiral quark--soliton 
model \cite{Kim96}, which were based on a different type of 
ultraviolet regularization (proper time regularization).
Furthermore, we observe a good agreement with the QCD sum rule 
calculations of Refs.\cite{Ioffe:1995aa, HeJi96}, and with the 
lattice estimates of Ref.\cite{Aoki:1997pi}.
\par
{\it Discussion of previous calculations of transversity 
distributions in the chiral quark--soliton model.} 
The transversity distributions 
have been computed by Gamberg {\it et al.}\ in a description of the
nucleon as a chiral soliton of the Nambu--Jona Lasinio 
model \cite{Gamberg:1998vg}, which is largely equivalent 
to the chiral quark--soliton model, and by Wakamatsu and 
Kubota \cite{WK98} in same approach as the one used here.
In the calculation of Ref.\cite{Gamberg:1998vg} 
only the contribution of the discrete bound--state level
was taken into account. It is known that in general this
simplification, which is not warranted by any parametric limit
of the model, leads to a number of inconsistencies 
({\it e.g.}\ the positivity of the isoscalar unpolarized antiquark 
distribution is violated), as has been discussed in detail in 
Refs.\cite{DPPPW96,DPPPW97}. In the transversity quark
distributions it so happens that the contributions from the Dirac 
continuum of quarks are numerically small, as shown by the 
results of our calculation in Fig.~\ref{fig_delta}. In the
antiquark distributions, however, the continuum contributions
are seen to be numerically important.
\par
In Ref.\cite{WK98} a calculation of the transversity distribution
was presented including also $1/N_c$--corrections
to the isovector distribution, $\delta u(x) - \delta d(x)$. As
was noted also in Ref.\cite{Gamberg:1998vg}, these
corrections are afflicted by ordering ambiguities of the collective
operators, and it is not clear if the ordering adopted in 
Ref.\cite{WK98} is correct. Short of a satisfactory answer to these
questions we have chosen to limit ourselves to the contributions
appearing in the lowest non-vanishing order of $1/N_c$. A detailed 
investigation of the $1/N_c$--corrections will be reported elsewhere.
\par
An important difference of our results to those of Ref.\cite{WK98}
concerns the ultraviolet regularization. In Ref.\cite{WK98} the 
Dirac continuum contributions to all distributions was subjected
to a Pauli--Villars subtraction, although our analysis shows that these 
quantities are ultraviolet finite and do not require regularization.
In contrast, we have left these distributions unregularized.
A more detailed comparison with the results of Ref.\cite{WK98} is,
unfortunately, not possible, since it is not clear how the
anomaly--type phenomenon discovered in our calculation 
(see Section~\ref{subsec_anomaly}) affects the results of 
Ref.\cite{WK98}, where the equivalence of summation over occupied
and non-occupied levels in the final results was not convincingly
established.
\subsection{Inequalities in the large $N_c$--limit}
\label{subsec_inequalities}
The formal limit of large $N_c$ allows one to derive certain positivity 
bounds for the transversity distributions, as well as generalizations 
of the Soffer inequalities, which are stronger than those which hold
in QCD at $N_c = 3$ \cite{Pobylitsa:2000tt}. It is interesting to check 
if these inequalities are satisfied by the results of the numerical 
calculations. 
\par
\begin{figure}[t]
\begin{center}
\begin{tabular}{cc}
\includegraphics[width=7.4cm,height=7.4cm]{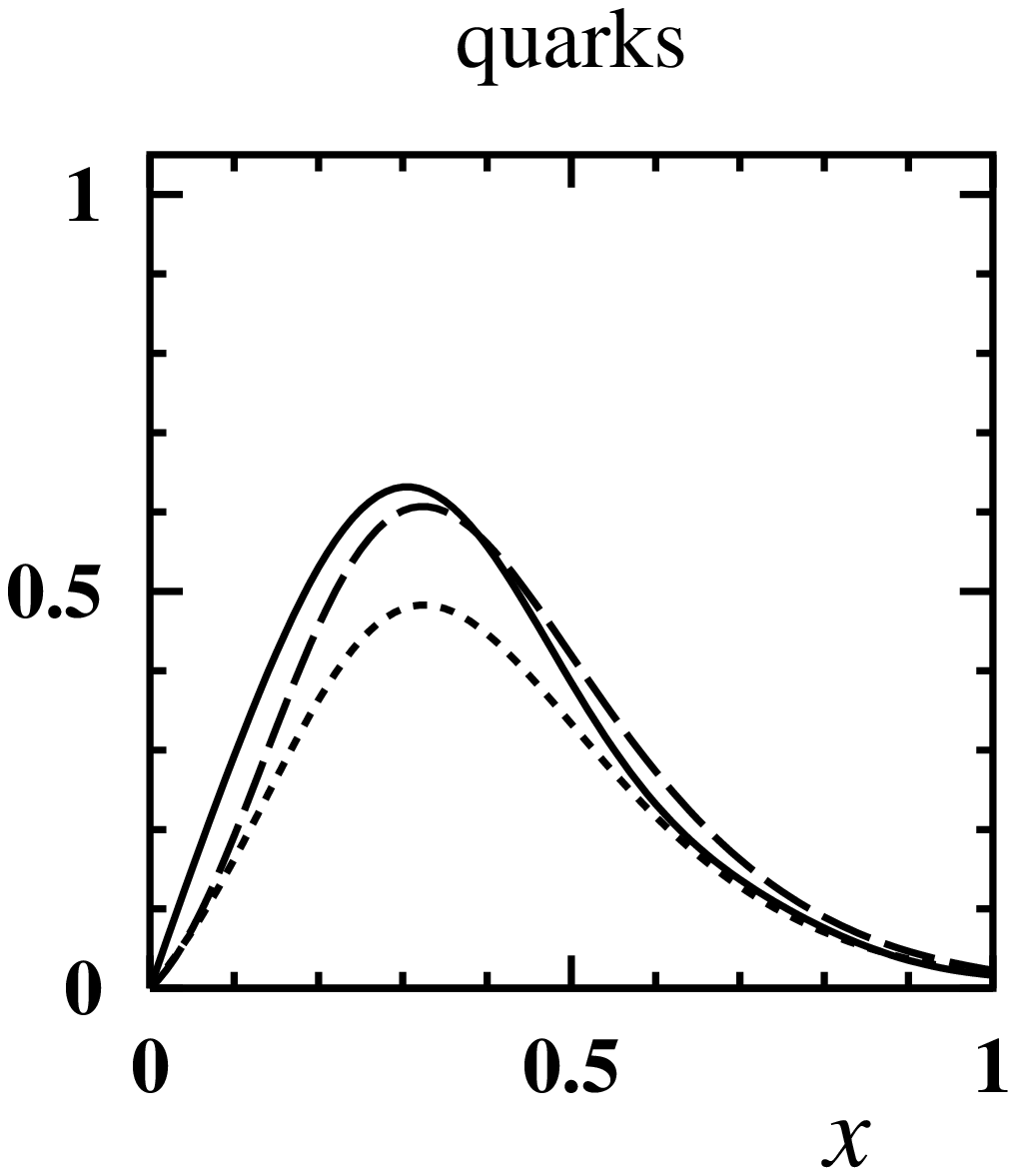}
&
\includegraphics[width=7.4cm,height=7.4cm]{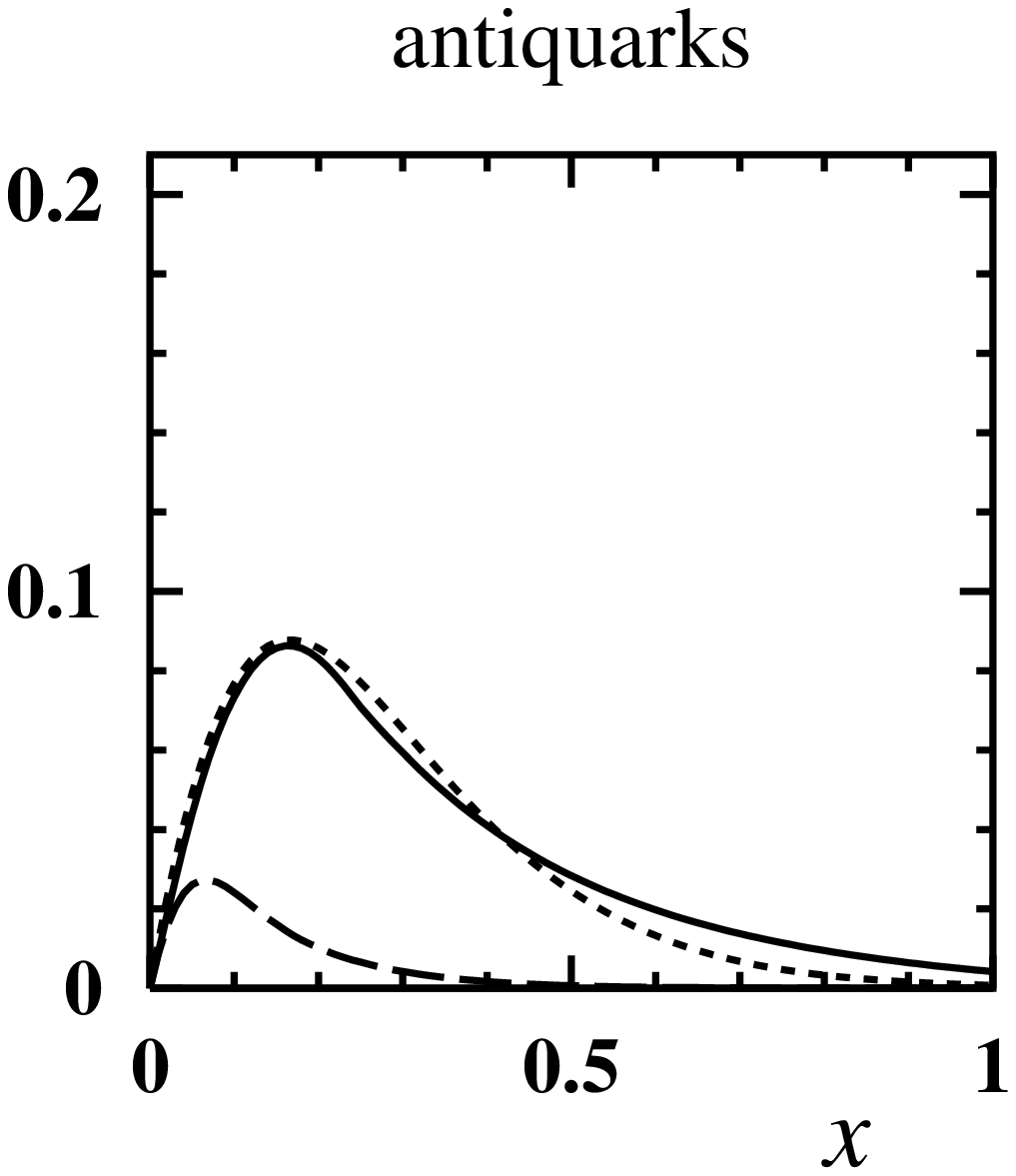}
\end{tabular}
\end{center}
\caption[]{\it 
The large--$N_c$ improved positivity bounds for quark distribution 
functions, Eqs.(\ref{E0}) and Eqs.(\ref{E1}), for the quark (left) 
and antiquark distributions (right).
\underline{Solid lines:}  $ x\, [u+d](x)/3$. 
\underline{Dashed lines:} $|x\,[\delta u -\delta d](x)|$.           
\underline{Dotted lines:} $ x\,[\Delta u -\Delta d](x) $.}
\label{xineq}
\vspace{10cm}
\end{figure}

In Ref.\cite{PP96} the following positivity constraints were 
derived in the large--$N_c$ limit of QCD:
\be
\frac{\,u(x)+d(x)\,}{3} &\ge&  |\Delta u(x)-\Delta d(x)| ,
\label{E0}
\\
\frac{\,u(x)+d(x)\,}{3} &\ge&  |\delta u(x)-\delta d(x)| ,
\label{E1}
\ee
and similarly for the corresponding antiquark distributions. 
In the chiral quark--soliton model the second inequality is trivially 
satisfied in the limit of large ultraviolet cutoff, since the
isoscalar unpolarized distribution is logarithmically divergent,
while the transversity distributions are finite. For finite cutoff,
however, the inequalities could in principle be violated by the 
ultraviolet regularization. In Fig.~\ref{xineq} we have plotted the 
results for both the L.H.S.\ and the R.H.S.\ in the chiral quark--soliton 
model. (The isoscalar unpolarized distribution has been taken from 
Ref.\cite{WG97}.) One sees that the numerical results respect the 
inequalities, both for the quark and antiquark distributions, to a very
good extent. The small numerical violation should be attributed to the
ultraviolet regularization. It is interesting that for the quarks
we observe saturation of the inequality for the transversity distribution, 
while the longitudinally polarized distribution is smaller than its bound,
while for antiquarks the situation is opposite: The 
longitudinally polarized distribution saturates the inequality, and
the transversity distribution falls short of its bound.
This qualitative behavior could be a useful guideline for parameterizing
the distributions in analysis of experimental data.
\par
\begin{figure}[t]
\begin{center}
\begin{tabular}{cc}
\includegraphics[width=7.4cm,height=7.4cm]{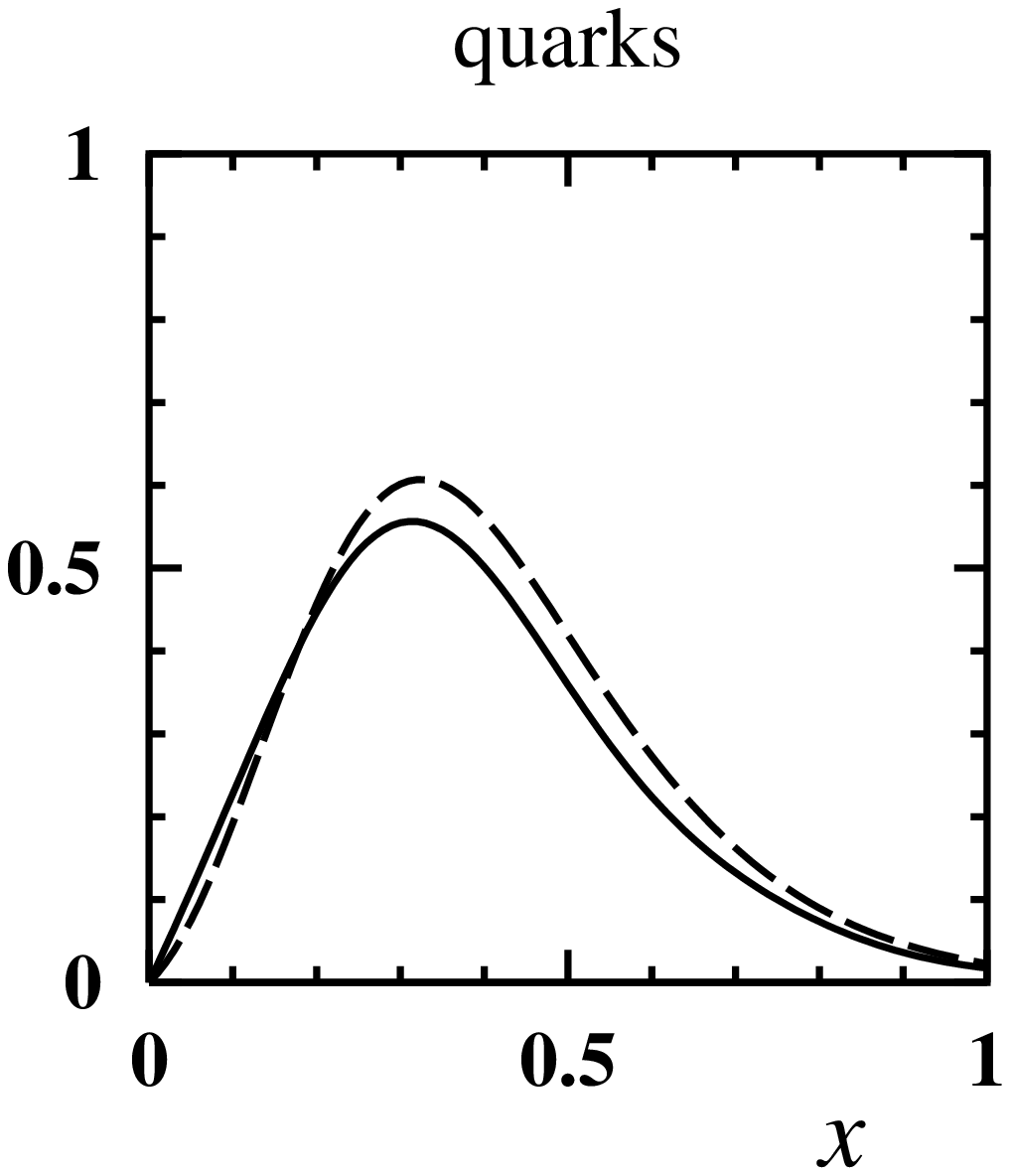}
&
\includegraphics[width=7.4cm,height=7.4cm]{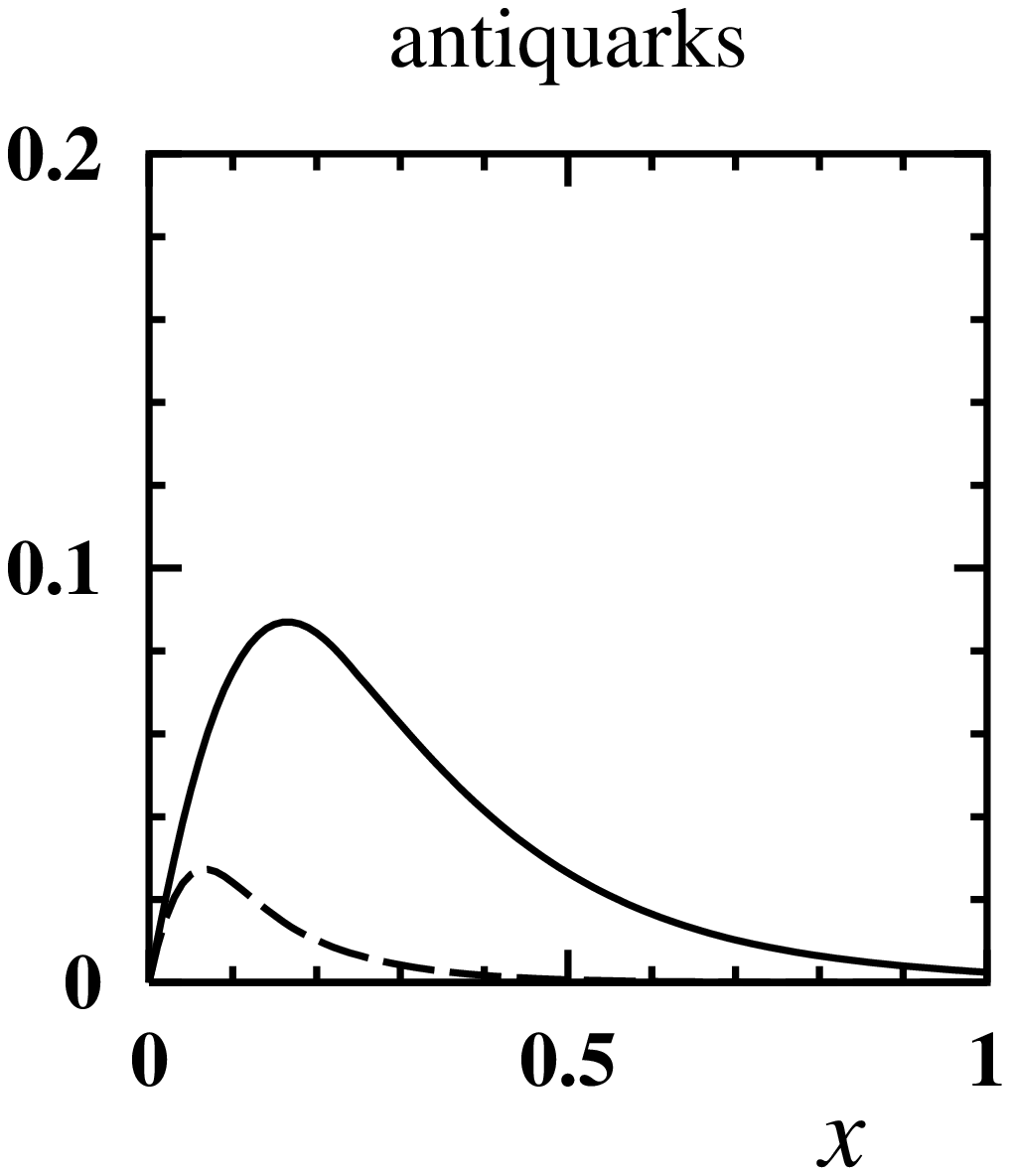}
\end{tabular}
\end{center}
\caption[]{\it
The large--$N_c$ improved Soffer bound for the quark (left) and 
antiquark (right) distributions. \underline{Dashed lines:} 
$|x\, [\delta u -\delta d](x)|$.
\underline{Solid lines:} The large $N_c$ Soffer bound 
$(x[u+d](x)/3+x\,[\Delta u -\Delta d](x))/2$.
The small violation is due to the ultraviolet regularization;
see the discussion in the main text.}
\label{xsoffer}
\vspace{10cm}
\end{figure}

Similar is the situation with the ``large $N_c$ version'' of the 
Soffer inequalities. In Ref.\cite{Pobylitsa:2000tt} the following
large--$N_c$ inequalities were proven:
\beq\label{E3}
 \frac{1}{2}\, \left[ 
\frac{u(x)+d(x)}{3}+ \left( \Delta u(x)-\Delta d(x) \right) \right]
\;\; \ge \;\; \left| \delta u(x)-\delta d(x) \right| \;.
\eeq
In Fig.~\ref{xsoffer} we plot the L.H.S.\ and the R.H.S.\ of this
inequality, as obtained in the model calculation in the large--$N_c$ limit.
The same caveats concerning the ultraviolet regularization apply as in
the case of the positivity conditions discussed above. As we can see, 
the model results satisfy the inequality within the expected accuracy.
\section{Transverse spin asymmetries in polarized 
Drell--Yan pair production}
\setcounter{equation}{0}
\label{sec_asymmetries}
With the numerical results for the transversity quark--and antiquark
distributions we can proceed to make predictions for observables which
would allow to extract these distributions from experiment. As explained
in the Introduction, the transversity distributions cannot be measured
in inclusive DIS at leading--twist level. In semi--inclusive DIS they
enter together with chirally odd fragmentation function, which, too, are 
essentially unknown quantities \cite{Jaffe97}. The cleanest way to measure
the transversity distribution at leading--twist level seems to be
Drell--Yan (DY) pair production in scattering of transversely polarized 
$pp$ or $p\bar p$. We shall therefore concentrate on this process here.
\par
The cross section for DY pair production is a function of
the center--of--mass energy of the incoming protons,
$s \!=\! (p_1 \!+\! p_2)^2$, and of the invariant mass of the produced
lepton pair, $M^2 \!=\! (k_1+k_2)^2$, 
which is equal to the virtuality of the exchanged photon and where
$k_{1/2}$ are the momenta of the detected leptons.
At the partonic level this process is described by the annihilation of a 
quark and an antiquark originating from the two protons, carrying, 
respectively, longitudinal momenta $x_1 p_1$
and $x_2 p_2$.\footnote{
	For questions concerning the reconstruction of the partonic initial 
	state from the event data, see {\it e.g.}\ 
	Ref.\cite{RHIC,RHICrecent}.}
The momentum fractions are given by $x_{1/2} \!= \! (Q^2 /s )^{1/2} e^{\pm y}$ 
with rapidity $y=\frac{1}{2}\ln\frac{p_1(k_1+k_2)}{p_2(k_1+k_2)}$.
We consider the case that the two protons are
transversely polarized relative to the beam direction.
In leading--order QCD the transverse spin asymmetry of the
DY cross section is given by
\be
A_{TT}^{pp} (y; s, M^2 ) &=& \frac{ \sum_f e_f^2 \; \delta q_f (x_1, M^2 )
\; \delta q_{\bar f} (x_2 , M^2) }{ \sum_f e_f^2 \; q_f (x_1, M^2 )
\; q_{\bar f} (x_2 , M^2) } ,
\label{A_TT}
\ee
where the sum runs over all species of light quarks and antiquarks
in the two nucleons, $f \! = \! \{ u, \bar u , d, \bar d,\,\dots \}$,
The relevant scale here for 
the parton distribution functions is the virtuality of the photon, $M^2$.
Note that this asymmetry is sensitive to the antiquark distributions. 
We neglect strange quark contributions here; since they always enter in 
the form of a product
of a strange quark with a strange antiquark distribution they can be 
expected to be very small.
\par
In the case of DY pair production in $p \bar p$ rather than $pp$ collisions,
using charge conjugation, $\delta q_{f / p} (x) \equiv 
\delta q_{\bar f / \bar p} (x)$ and $q_{f / p} (x) \equiv 
q_{\bar f / \bar p} (x)$, the above expression changes to 
\be
A_{TT}^{p \bar p} (y; s, M^2 ) &=& 
\frac{ \sum_f e_f^2 \; \delta q_f (x_1, M^2 )
\; \delta q_f (x_2 , M^2) }{ \sum_f e_f^2 \; q_f (x_1, M^2 )
\; q_f (x_2 , M^2) } .
\label{A_TT_ppbar}
\ee
\par
The transverse spin asymmetry, Eqs.(\ref{A_TT}) and (\ref{A_TT_ppbar}), 
requires the quark-- and
antiquark distributions of the individual quark flavors 
$\delta u (x), \delta d (x), \delta \bar u (x)$ and $\delta \bar d (x)$.
In our approach, based on the large--$N_c$ limit, we have computed
the isovector distribution [$\delta u (x) - \delta d (x),  
\delta \bar u (x) - \delta \bar d (x)$], which appears in
leading order of the $1/N_c$--expansion, and the isoscalar 
[$\delta u (x) + \delta d (x),  
\delta \bar u (x) + \delta \bar d (x)$], which appears in next--to--leading
order, both in the lowest non-vanishing order of $1/N_c$. 
From these results one should not, strictly speaking, recover the distribution
of the individual flavors by adding and subtracting the isovector and 
isosinglet combination, since there can be $1/N_c$--corrections to the 
isovector distribution, of the same order in $1/N_c$ as the isosinglet 
distribution, which are not included. 
Such $1/N_c$--corrections were studied in Ref.\cite{WK98}, however we did not 
compute them here but rather take a pragmatic stand. We directly reconstruct 
the individual flavor distributions from the results of the model calculations.
For the quark transversity distributions we take the model result for 
the ratios of transversity to longitudinally polarized distributions,
Eqs.(\ref{ratio_isovector}) and (\ref{ratio_isoscalar}), together with the
GRSV 95 LO distribution for the longitudinally polarized distributions.
For the antiquark transversity distributions we neglect the contribution
of the isoscalar distribution as it is much smaller than the isovector one
(see Figs.\ref{fig_delta} and \ref{fig_xdelta}).
\par
In order to get an impression of the dependence of the prediction
for the transverse spin asymmetries on the choice of transversity
distributions we compare the above results with the asymmetries 
calculated under the assumption that $\delta q (x) \equiv \Delta q (x), 
\delta \bar q (x) \equiv \Delta \bar q (x)$, using again the GRSV 95 
parameterization for $\Delta q (x)$ and $\Delta \bar q (x)$. (This choice,
which is consistent with the Soffer inequalities, has frequently been 
made in the literature \cite{Ji92,Barone97}.) 
\par
\begin{figure}[t]
\begin{center}
\begin{tabular}{cc}
\includegraphics[width=7.4cm,height=7.4cm]{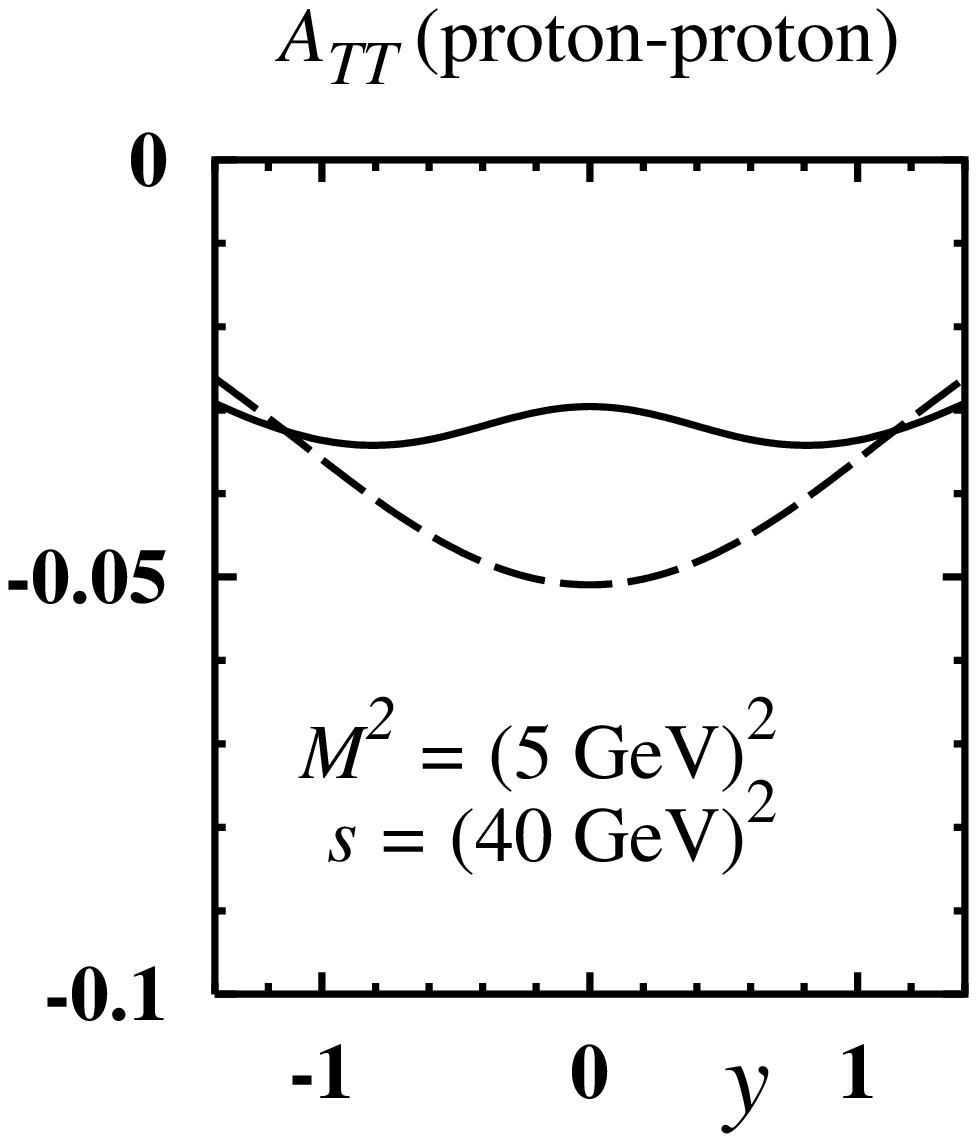}
&
\includegraphics[width=7.4cm,height=7.4cm]{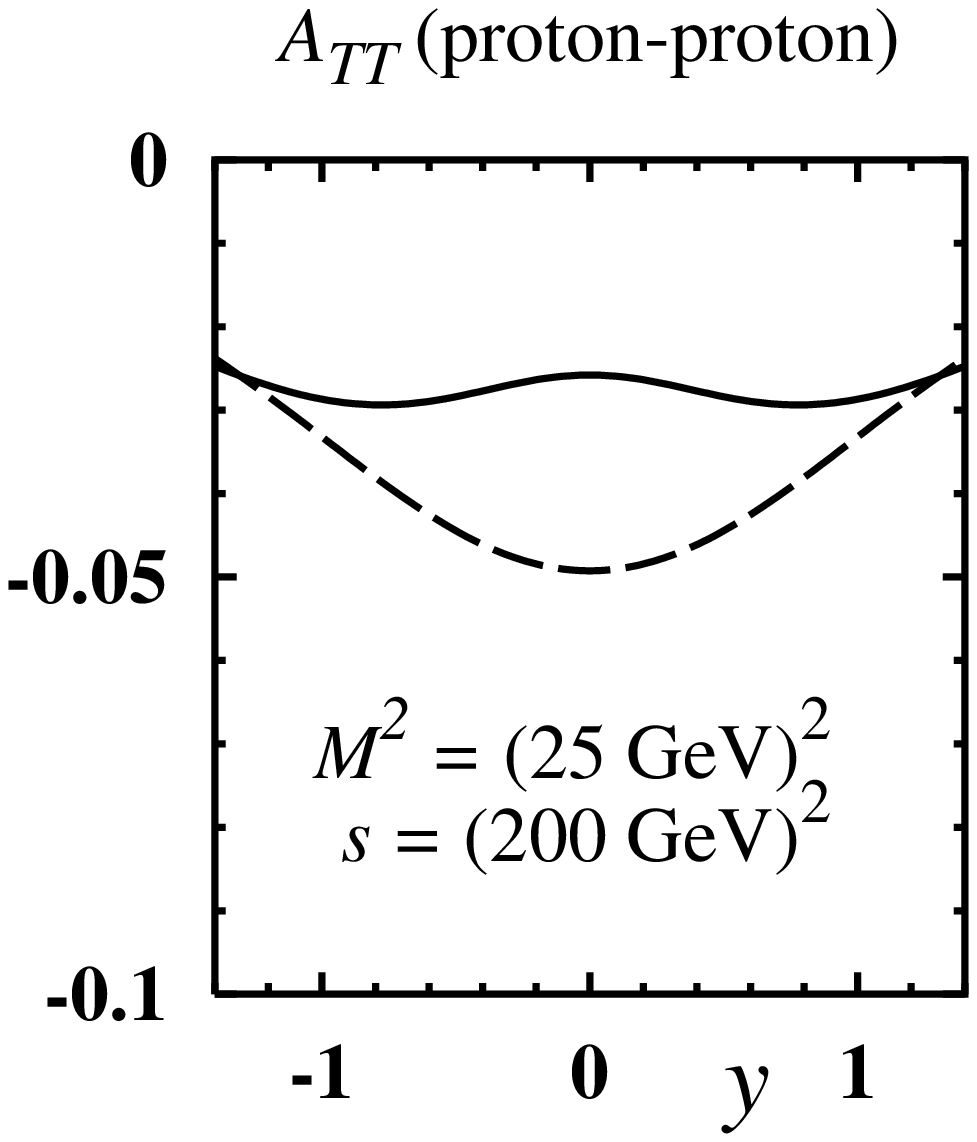}
\end{tabular}
\end{center}
\caption[]{\it The transverse spin asymmetry, $A_{TT}$, 
Eq.(\ref{A_TT}), in collisions of transversely polarized protons, 
in two different kinematical regions: 
$s = (40\,{\rm GeV})^2, M^2 = (5 \,{\rm GeV})^2$ (left),
and $s = (500 \,{\rm GeV})^2, M^2 = (25 \,{\rm GeV})^2$ (right).
\underline{Solid lines:} Asymmetries calculated with the quark-- and 
antiquark distributions computed in the chiral quark--soliton 
model, {\it cf.}\ Fig.\ref{fig_xdelta}. For the quark distributions
we used the calculated ratios of transversity to longitudinally polarized 
distributions, Eqs.(\ref{ratio_isovector}) and (\ref{ratio_isoscalar}),
together with the GRSV95 parameterizations {\rm \cite{GRSV96}} for 
$[\Delta u - \Delta d](x)$ and $[\Delta u + \Delta d](x)$. 
\underline{Dashed lines:} Asymmetries obtained assuming
that $\delta q (x) \equiv \Delta q (x)$ and 
$\delta \bar q (x) \equiv \Delta \bar q (x)$ ($q = u, d$), using the GRSV95 
parameterizations {\rm \cite{GRSV96}} 
for $\Delta q(x)$ and $\Delta \bar q(x)$.}
\label{fig_att}
\vspace{10cm}
\end{figure}

The transverse spin asymmetries obtained with the two ``scenarios'' for the 
transversity distributions are shown in Fig.~\ref{fig_att}.
One sees that the differences in the $pp$ asymmetries are quite sizable,
in particular in the region of small rapidities. Note, however, that
these observables depend very sensitively on the small antiquark 
distributions, which may be affected by $1/N_c$--corrections.
Even greater differences are seen in the asymmetries for $p\bar p$
reactions. Here the asymmetries are dominated by the products of
quark distributions in the proton, while the contributions from the 
antiquark distributions are negligible. The differences between 
the asymmetries calculated with our model distributions and those 
with $\delta q (x) \equiv \Delta q(x)$ essentially reflect the 
numerical enhancement of the transversity quark distributions over the
longitudinally polarized ones found in the model calculations, {\it cf.}\
Eqs.(\ref{ratio_isovector}) and (\ref{ratio_isoscalar}).
To summarize, we find that our model distributions
result in significant deviations from what is obtained with that
approximation. Unfortunately, recent studies, taking into account
the limited detector acceptance, suggest that measurements at RHIC are 
unlikely to be able to discriminate between the two 
scenarios \cite{Martin:1999eu}.
\section{Conclusions}
\setcounter{equation}{0}
In this paper we have studied the transversity quark--and antiquark
distributions in a dynamical model of the nucleon based on the 
large--$N_c$ limit. Let us briefly summarize the main qualitative 
conclusions.
\par
Comparing the transversity and longitudinally polarized distributions,
we found that, generally speaking, the quark distributions 
(both isovector and isoscalar) are comparable in magnitude. The
corresponding antiquark distributions, on the contrary, are very different. 
Appealing to the gradient expansion, which, as we saw, gives a realistic 
numerical description of the antiquark distributions, we were able
to explain the smallness of the isovector transversity antiquark 
distribution relative to its longitudinally polarized counterpart
on grounds of the constraints imposed on the low--energy effective 
dynamics by chiral symmetry. A measurement of the transversity antiquark
distributions would thus be a sensitive test of the role of chiral 
symmetry in determining the parton distributions of the nucleon 
at a low scale.\footnote{Also, it would be interesting to see 
if the the differences between the longitudinally and transversity 
antiquark distributions can qualitatively be understood in the popular
``pion cloud'' picture, which is widely used to explain the flavor
asymmetry of the unpolarized antiquark distributions, $\bar u - \bar d$.
We stress that that model is not related to the large--$N_c$ approach
employed in this work, see Refs.\cite{Koepf:1996yh,Dressler:2000zg} 
for a critical discussion.
The issue of polarization of the antiquark distributions in the
``pion cloud'' model has recently been discussed in 
Ref.\cite{Dressler:2000zg}.} 
\par
The strong differences between the longitudinally and transversity 
antiquark distributions should be taken into account when making predictions
for observables sensitive to the antiquark distributions. Our estimates show
that these differences have a noticeable effect {\it e.g.}\ on the spin 
asymmetries in polarized Drell--Yan pair production. It remains to be seen
if these asymmetries can be measured to an accuracy that would allow one
to discriminate between the different predictions \cite{Martin:1999eu}.
\par
Also, we have verified that the quark/antiquark distributions obtained
in the chiral quark--soliton model satisfy the Soffer inequalities,
as well as the large--$N_c$ inequalities derived in 
Ref.\cite{Pobylitsa:2000tt}. Since these relations could in principle
be violated by the ultraviolet cutoff required in the model calculation,
their fulfillment should be seen as another piece of evidence in favor of
the Pauli--Villars regularization scheme adopted 
here \cite{DPPPW96,DPPPW97,WG97}.
\par
We have noted a curious anomaly--type phenomenon in the tensor 
charge and the transversity distribution functions. Here we have described
this phenomenon using the language of sums over quark single--particle
in the background pion field specific to our large--$N_c$ picture
of the nucleon. It is possible, however, that the anomaly described
here has meaning also outside of the context of this model. To clarify
this one should see if the observations made here could be stated in
a more general, field--theoretic language. The analogy with the
Fujikawa formulation of the $U(1)$ anomaly, whose field--theoretical
description is well--known, should be a useful guideline.

\section*{Acknowledgements}
 We thank T.~Watabe for his help during the initial stages of this work,
 B.~Dressler for kindly providing us with a computer code for the
 leading--order evolution of the transversity distributions,
 and A.~V.~Efremov for inspiring conversations.
\newpage
\appendix
%
%
%
\renewcommand{\theequation}{\Alph{section}.\arabic{equation}}
\section{Anomaly--type phenomenon in the transversity distribution function}
\label{app_anomaly}
\setcounter{equation}{0}
In the calculation of the isovector transversity distribution function in the
effective chiral theory we encounter an anomaly--type phenomenon:
The distributions obtained by summing the contributions
of occupied and non-occupied quark levels in the soliton are different
if the soliton field does not fall off faster than $1/r^2$ at large 
distances. In Section~\ref{subsec_anomaly} we computed the anomalous
difference in the first moment of the isovector distribution, {\it i.e.}, 
the isovector tensor charge. Using a somewhat different approach
one can easily compute the anomalous difference in the $x$--dependent 
distribution function. Since the anomaly phenomenon in the transversity
distribution is of principal theoretical interest, as well as of practical 
importance for the numerical calculations (see Section~\ref{subsec_numerical}),we derive here the expression for the anomalous difference of the
isovector distribution function.
\par
The representation of the isovector transversity distribution as a
sum over occupied quark levels, Eq.(\ref{isovector_occ}), can be written
as an integral over a continuous energy variable, $\omega$, in the 
form\footnote{Throughout this section it will be understood that
$\delta u(x)$ {\it etc.}\ denotes the quark distribution for $x > 0$, 
and minus the antiquark distribution at $x < 0$, {\it cf.}\ 
Eq.(\ref{delta_qbar_def}).}
\beq
\left[ \delta u(x) - \delta d (x) \right] _{\rm occup.} 
\; = \; \int\limits_{-\infty }^{E_{lev}+0}d\omega \; \rho (\omega )
\eeq
where the integrand $\rho (\omega )$ is defined by the functional
trace of the quark ``density of states'', $\delta (\omega - H)$, with the 
relevant single--particle operator
\beq
\rho (\omega ) \;\; \equiv \;\; \frac{2 N_cM_N}{3}
\mbox{Sp} \left[ \delta
(\omega -H)\delta (\omega +P^3-xM_N)\frac{1 + \gamma ^0\gamma ^3}{2}
\gamma ^5\gamma^1\tau^1 \right] \; .
\eeq
Here, $P^i$ denotes the single--particle three--momentum operator,
and $H$ the Dirac Hamiltonian, Eq.(\ref{H}).
Similarly, the representation of the distribution as a
sum over non-occupied quark levels, Eq.(\ref{isovector_nonocc}), can be 
written in the form
\beq
\left[ \delta u(x) - \delta d (x) \right] _{\rm non-occup.} 
\;\; = \;\; -\int\limits_{E_{lev}+0}^{\infty} d\omega \; \rho (\omega ) .
\eeq
The difference between the two representations is thus given as an 
integral over {\it all} energies:
\beq
\left[ \delta u(x) - \delta d (x) \right] _{\rm occup.} 
- \left[ \delta u(x) - \delta d (x) \right] _{\rm non-occup.} 
\;\; = \;\; \int\limits_{-\omega _0}^{\omega _0} 
d\omega \; \rho (\omega ) .
\label{difference_x_starting}
\eeq
We have introduced here an explicit energy cutoff, $\omega_0$, in order to 
specify the limiting procedure leading to the anomaly. In order to 
analyze the difference, Eq.(\ref{difference_x_starting}), in the limit of
large $\omega_0$ we write the delta function of $\omega - H$ as the
imaginary part of the quark propagator \cite{DPPPW97}:
\be
\lefteqn{
\left[ \delta u(x) - \delta d (x) \right] _{\rm occup.} 
- \left[ \delta u(x) - \delta d (x) \right] _{\rm non-occup.} } 
\nonumber && \\
&=& 
\frac{4 N_c M_N}{3} 
\mbox{Im} \int\limits_{-\omega _0}^{\omega _0}\frac{d\omega }{%
2\pi }\mbox{Sp}\left[ \frac 1{H-\omega -i0}\delta (\omega
+P^3-xM_N)\frac{1+\gamma ^0\gamma ^3}{2}\gamma _5\gamma ^1\tau^1\right] \; .
\nonumber \\
\label{difference_x_ready}
\ee
We now proceed as in the case of the tensor charge in 
Section~\ref{subsec_anomaly}, and expand the integrand for large $\omega$.
This can be done by writing the quark propagator in the form
\beq
\frac 1{H-\omega -i0} \;\; = \;\;
\frac {H + \omega}{H^2-\omega ^2-i0\mbox{sign}\omega } ,
\eeq
and substituting the formal series expansion of this expression in powers 
of derivatives of the pion field, Eq.(\ref{propagator_series}), 
collecting all terms 
contributing in a certain order of $1/\omega$. In our case 
the leading contribution in $1/\omega$ comes from term with $n = 1$
in Eq.(\ref{propagator_series}); the term with $n = 0$ cannot 
produce a $\gamma^1$ Dirac matrix needed to compensate the 
$\gamma^1$ matrix in the operator in Eq.(\ref{difference_x_ready}).
Taking into account also $\gamma_5$--parity one finds that the
leading contribution at large $\omega$ is given by
\be
\lefteqn{
\left[ \delta u(x) - \delta d (x)\right] _{\rm occup.} 
- \left[ \delta u(x) - \delta d (x)\right] _{\rm non-occup.} } 
\nonumber && \\
&=& 
\frac{4 N_c M_N}{3} \mbox{Im}\int\limits_{-\omega _0}^{\omega _0}
\frac{d\omega}{2\pi} 
\mbox{Sp}\left\{ \frac{(\omega -i\gamma ^0\gamma ^k\partial _k)}{\left[
-\partial ^2+M^2-\omega ^2-i0\mbox{sign}\omega \right] ^2}
\delta (\omega +P^3-xM_N) 
\right.
\nonumber \\
&& \left. \times \frac{1+\gamma ^0\gamma ^3}{2}
\gamma _5\gamma ^1\tau^1 \left[ -iM(\gamma
^k\partial _kU_{\rm cl}^{\gamma _5})\right] \right\} \;.
\ee
We can evaluate the functional trace in the basis of 
momentum eigenstates, $\mbox{Sp}\left[ \ldots \right] = 
\int d^3 p / (2 \pi)^3 \, \langle p | \ldots | p \rangle$, inserting 
complete sets of intermediate position eigenstates in which the pion 
field is diagonal. In addition, we have to take the trace over Dirac
and flavor indices. In this way we obtain
\be
\lefteqn{
\left[ \delta u(x) - \delta d (x)\right] _{\rm occup.} 
- \left[ \delta u(x) - \delta d (x)\right] _{\rm non-occup.} } 
\nonumber && \\
&=& 
\frac{4 N_c M_N}{3} \mbox{Im}\int\limits_{-\omega _0}^{\omega _0}
\frac{d\omega }{2\pi} \int\frac{d^3p}{(2\pi )^3} 
\mbox{Tr}_{\rm Dirac} \left\{ \frac{(\omega +\gamma ^0\gamma ^kp^k)\delta
(\omega +p^3-xM_N)}{\left[ |{\bf p}|^2 + M^2
-\omega^2 - i0\mbox{sign}\omega \right]^2}\right. 
\nonumber \\
&& \times \left. \frac{1+\gamma ^0\gamma ^3}{2} \gamma _5\gamma ^1
\int d^3 x \, \mbox{Tr}_{\rm flavor} \left[ -iM \gamma^k
\partial_k U_{\rm cl}^{\gamma _5}({\bf x}) \tau^1 \right] \right\} 
\\
&=& \frac{16 N_c M_N M }{3} \, 
\int d^3x \mbox{Tr}_{\rm flavor}
\left\{ i \tau^1 \partial_1 \left[ U_{\rm cl}({\bf x}) - U_{\rm cl}^\dagger ({\bf x}) 
\right] \right\}
\nonumber \\
&& \times \mbox{Im}
\int\limits_{-\omega _0}^{\omega _0}\frac{d\omega }{2\pi}
\int \frac{d^3p}{(2\pi )^3}\frac{(\omega +p^3)\delta (\omega +p^3-xM_N)}
{\left[ |\mathbf{p}|^2+M^2-\omega ^2-i0\mbox{sign}\omega \right] ^2} \; . 
\label{difference_x_integral}
\ee
It remains to compute the integral over the energy, $\omega$, and 
the momentum, $p$, in the last expression. Instead of computing the
integral first and then taking its imaginary part it is convenient
to take the imaginary part ``under the integral'', substituting
\be
\lefteqn{
\mbox{Im} \left[ (\omega +p^3)\delta (\omega +p^3-xM_N)
\frac{1}{\left( |\mathbf{p}|^2+M^2-\omega ^2-i0\mbox{sign}\omega \right)^2}
\right] } &&
\nonumber \\
&\rightarrow & (\omega +p^3)\delta (\omega +p^3-xM_N)
\pi (-\mbox{sign}\omega )\delta ^{\prime }(|\mathbf{p}|^2+M^2-\omega ^2) 
\nonumber \\
&=& x M_N \delta (\omega +p^3-xM_N)
\pi (-\mbox{sign}\omega ) \delta ^{\prime }\left[ 
-xM_N(\omega -p^3) + |{\bf p}^{\perp }|^2 + M^2 \right] .
\ee
In the last step here we have made use of the condition imposed by the
first delta function in order to simplify the integrand. The integral over
$p^3$ can be taken using up the first delta function:
\be
\lefteqn{
\int\limits_{-\omega _0}^{\omega _0}\frac{d\omega }{2\pi}
\int \frac{d^3p}{(2\pi )^3} \; \mbox{Im} \left[ \ldots \right] }
&& \nonumber \\
&=&
\int\limits_{-\omega _0}^{\omega _0}\frac{d\omega }{2\pi }
\int \frac{d^2p^{\perp }}{(2\pi )^2} x M_N 
(-\mbox{sign}\omega )\delta ^{\prime }\left[ -xM_N(2\omega -xM_N)
 + |{\bf p}^{\perp }|^2 + M^2\right] .
\ee
The integral over $\omega $ requires some care. We have
\be
&& 
\int\limits_{-\omega _0}^{\omega _0}\frac{d\omega }{2\pi }(-\mbox{sign}
\omega )\delta ^{\prime }\left[ -xM_N(2\omega -xM_N)+|\mathbf{p}^{\perp
}|^2+M^2\right] 
\\
&=& 
\left( -\int\limits_0^{\omega _0} + \int\limits_{-\omega _0}^0 \right)
\frac{d\omega }{2\pi } \; \delta ^{\prime }\left[
-xM_N(2\omega -xM_N)+|\mathbf{p}^{\perp }|^2+M^2\right] 
\nonumber \\
&=&
\frac{1}{2xM_N}\frac 1{2\pi }\delta \left[ -xM_N(2\omega _0-xM_N)+
|\mathbf{p}^{\perp }|^2+M^2\right] 
\; + \; (\omega_0 \rightarrow -\omega_0 ).
\ee
The remaining integral over the transverse component of $p$ can easily
be performed after replacing 
$\int\frac{d^2 p^\perp}{(2 \pi )^2} 
 \rightarrow  \int\frac{d |{\bf p}^\perp |^2}{4\pi}$.
It gives rise to a sum of two step functions, which in the limit 
of large cutoff, $\omega_0 \rightarrow \infty$, are non-zero for
values of $x$ in the range
\beq
-2\omega_0/M_N < x < 2\omega_0 / M_N .
\eeq
Collecting everything we obtain from Eq.(\ref{difference_x_integral})
\be
\lefteqn{
\left[ \delta u(x) - \delta d (x)\right] _{\rm occup.}
-\left[ \delta u(x) - \delta d (x)\right]_{\rm non-occup.} } && \nonumber \\
&=& 
\frac{N_cM_NM}{6 \pi^2} 
\theta \left( \frac{-2\omega_0}{M_N} < x < \frac{2\omega_0}{M_N} 
\right)
\int d^3x\mbox{Tr}_{\rm flavor}\left[ i\tau^1
\partial_1(U_{\rm cl} - U_{\rm cl}^{+})\right] .
\label{anomaly_x_res}
\ee
Eq.(\ref{anomaly_x_res}) represents
the result for the ``anomalous difference'' between the sums over
occupied and non-occupied levels.
\par
As in the case of the tensor charge, Eq.(\ref{A_linear_divergence}), 
the anomalous difference of the distribution functions,
Eqs.(\ref{anomaly_x_res}),
is given by a total derivative of the pion field, which is non-zero
only if the field drops like $1/r^2$ at large distances, 
$r \rightarrow \infty$. In this case the unitary matrix at large
$r$ takes the form
\beq
U_{\rm cl} \;\; = \;\; 1 + \frac{\kappa}{r^2} i(n^k\tau^k) ,
\label{U_large_r}
\eeq
where the constant $\kappa$ is related to the nucleon 
isovector axial coupling constant in the chiral limit
by Eq.(\ref{g_A_from_kappa}). The integral of the total
derivative of the pion field then 
becomes:\footnote{
Instead of converting the integral
of the total derivative of the pion field into a surface 
integral at $r = \infty$, one also may directly differentiate 
the asymptotic form of the pion field, Eq.(\ref{U_large_r}),
and compute the volume integral. In this case one would find 
that the derivative of the $1/r^2$--term in Eq.(\ref{U_large_r})
gives rise to a delta function at $r = 0$, whose volume integral
reproduces Eq.(\ref{integral_total_derivative}).}
\be
&& \!\!\!\!\!\!\!\!
\int d^3 x\mbox{Tr}_{\rm flavor}\left[ i\tau^1
\partial_1(U_{\rm cl} - U_{\rm cl}^\dagger )\right]
= \frac{1}{3} \int d^3 x \mbox{Tr}_{\rm flavor} 
\left[ \sum\limits_{k=1}^3i\tau^k\partial_k (U_{\rm cl}-U_{\rm cl}^\dagger)
\right] 
\nonumber \\
&& 
= \lim_{r\rightarrow \infty } \frac{4\pi r^2}{3} 
\mbox{Tr}_{\rm flavor}
\left[ i\sum\limits_{k=1}^3n^k\tau^k(U_{\rm cl}-U_{\rm cl}^\dagger )\right]
= -\frac{4\pi \kappa}{3} \;\; = \;\; \frac{2 g_A}{9 F_\pi ^2} \,.
\label{integral_total_derivative}
\ee
The anomalous difference is thus given by
\be
\lefteqn{
\left[ \delta u(x) - \delta d (x) \right] _{\rm occup.}
-\left[ \delta u(x) - \delta d (x) \right]_{\rm non-occup.} 
} && \nonumber \\
&=& 
\frac{1}{12 \pi ^2}\frac{g_A}{F_\pi ^2}N_cM_NM\,\theta \left( -\frac{2\omega
_0}{M_N}<x<\frac{2\omega _0}{M_N}\right) \, .
\ee
The support of this function increases with the energy cutoff, $\omega_0$.
A similar phenomenon was observed in the anomalous difference in the
isoscalar unpolarized distribution in Ref.\cite{DPPPW97}. The moments
of this function (with respect to the variable $x$) thus exhibit
power divergences for $\omega_0 \rightarrow \infty$. In particular,
the first moment is linearly divergent:\footnote{In the large--$N_c$
limit the support of the parton distributions is not limited to
the $-1 < x < 1$, so we must integrate over the entire real axis
in order to recover the tensor charge as defined by the matrix element
of the local operator, Eq.(\ref{tensor_def}).}
\beq
\int\limits_{-\infty }^\infty dx \left\{
\left[ \delta u(x) - \delta d (x) \right]_{\rm occup.}
-\left[ \delta u(x) - \delta d (x)\right]_{\rm non-occup.} \right\}
\;\; = \;\;
\frac{N_c M g_A \omega _0}{3\pi ^2 F_\pi ^2} .
\label{anomaly_x_first_moment}
\eeq
This is in qualitative agreement with the behavior of the tensor charge 
in the limit of large energy cutoff derived in 
Section~\ref{subsec_anomaly}. Note that the coefficient
of the linear divergence depends on the details of the ultraviolet
cutoff used, so one should not expect to find the same 
coefficients in Eq.(\ref{anomaly_x_first_moment}) 
and Eq.(\ref{A_linear_divergence}).
%

%
%
%
%
\newpage
%
%
\newpage
%
%
%
%
%
%
\newpage
%
%
\newpage
\begin{figure}[t]
\begin{center}
\begin{tabular}{cc}
\includegraphics[width=7.4cm,height=7.4cm]{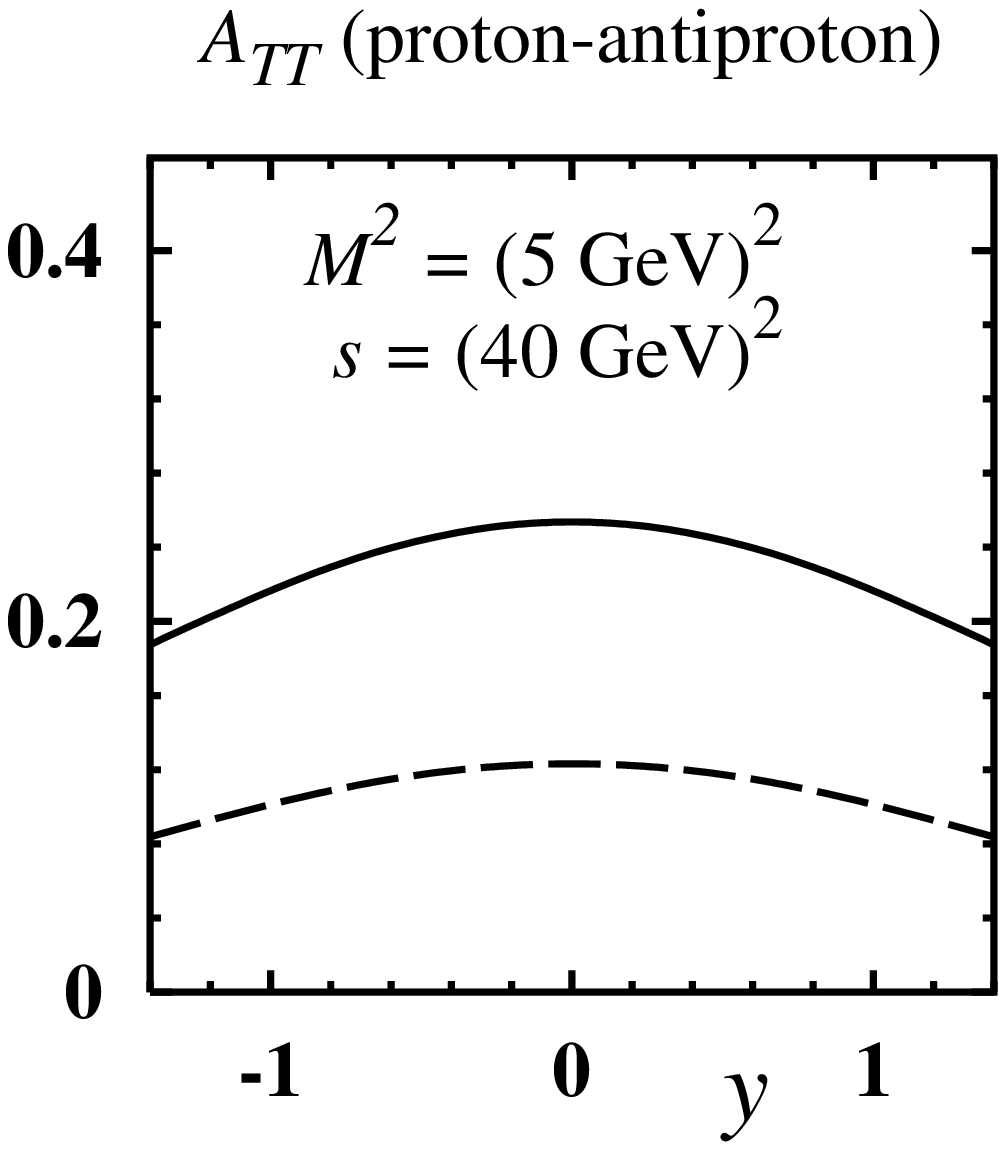}
&
\includegraphics[width=7.4cm,height=7.4cm]{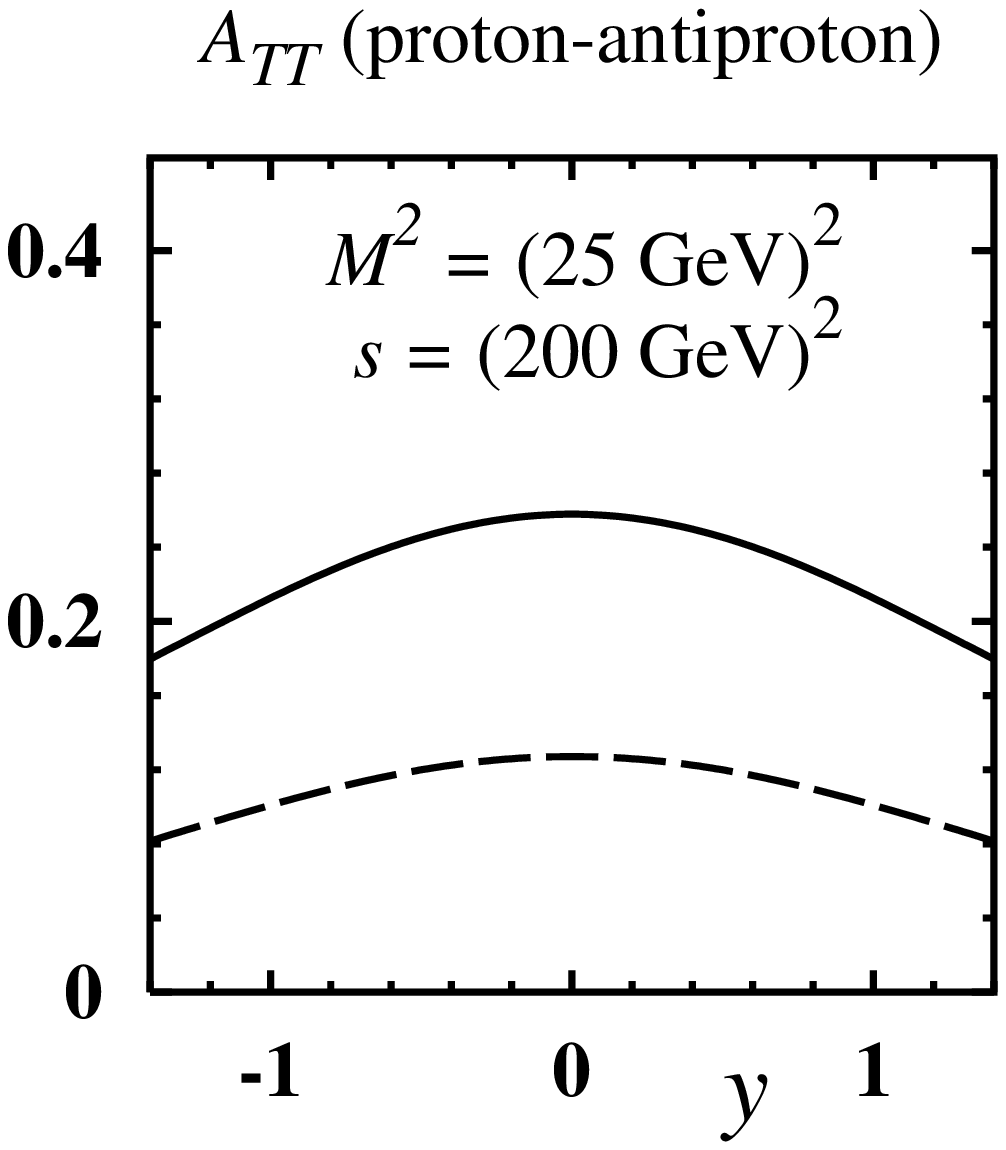}
\end{tabular}
\end{center}
\caption[]{\it The transverse spin asymmetry, $A_{TT}^{p\bar p}$, 
Eq.(\ref{A_TT_ppbar}), in collisions of transversely polarized 
protons and antiprotons. 
The solid and dashed lines correspond to the cases
described in Fig.\ref{fig_att}.}
\label{fig_att_ppbar}
\vspace{10cm}
\end{figure}
%
%
%

\newpage
\begin{thebibliography}{99}
%
\bibitem{Jaffe97}
R.L.\ Jaffe, Proceedings of ``Deep inelastic scattering off 
polarized targets, Physics with polarized protons at HERA'', 
Hamburg/Zeuthen 1997, p. 167-180, Report MIT-CTP-2685, hep-ph/9710465.
%
\bibitem{RS79} J.P.\ Ralston and D.E.\ Soper, Nucl.\ Phys.\ {\bf B 152}
(1979).
%
\bibitem{GRSV96} M.\ Gl\"uck, E. Reya, M. Stratmann, and W. Vogelsang,
Phys.\ Rev.\ {\bf D 53} (1996) 4775.
%
\bibitem{GRV95}
M.\ Gl\"uck, E.\ Reya and A.\ Vogt, Z.\ Phys.\ {\bf C 67} (1995) 433.
%
\bibitem{MRS98}
A.D.\ Martin, R.G.\ Roberts, W.J.\ Stirling, and 
R.S.\ Thorne, Eur.\ Phys.\ J.\ {\bf C 4} (1998) 463.
%
\bibitem{CTEQ99}
The CTEQ Collaboration:
H.L.\ Lai, J.\ Huston, S.\ Kuhlmann, J.\ Morfin, F.\ Olness, 
J.F.\ Owens, J.\ Pumplin, W.K.\ Tung, 
Report MSU-HEP/903100, hep-ph/9903282.
%
\bibitem{JJ91} R.L.\ Jaffe and X.--D.\ Ji, 
Phys.\ Rev.\ Lett.\ {\bf 67} (1991) 552;
Nucl.\ Phys.\ {\bf B 375} (1992) 527.
%
\bibitem{Ji92} X.-D.\ Ji, Phys.\ Lett.\ {\bf B 284} (1992) 137.
%
\bibitem{JaffeSaito96}
R.L.\ Jaffe and N.\ Saito, Phys.\ Lett.\ {\bf B 382} (1996) 165.
%
%
\bibitem{Collins}
J.Collins, Nucl.\ Phys.\ {\bf B 396} (1993) 161; \\
X.\ Artru and J.\ Collins, Z.\ Phys.\ {\bf C 69} (1996) 277.
%
%
\bibitem{Kotzinian:1997wt}
A.~M.~Kotzinian and P.~J.~Mulders,
Phys.\ Lett.\  {\bf B406} (1997) 373, hep-ph/9701330.
%
%
\bibitem{Boer98}
D.~Boer and P.J.~Mulders, Phys.\ Rev.\ {\bf D 57} (1998) 5780; \\
D.~Boer, R.~Jakob, and P.J.~Mulders, Phys.\ Lett.\ {\bf B 424} (1998) 143; \\
D.~Boer and R.~Tangerman, Phys.\ Lett.\ {\bf B 381} (1996) 305; \\
P.J.~Mulders and R.~Tangerman,  Nucl.\ Phys.\ {\bf B 461} (1995) 234.
%
%
\bibitem{Efremov:1999mv}
A.~V.~Efremov,
hep-ph/0001214.
%
%
\bibitem{Korotkov:1999jx}
V.~A.~Korotkov, W.~D.~Nowak and K.~A.~Oganesian,
hep-ph/0002268.
%
%
\bibitem{Kotzinian:1999dy}
A.~M.~Kotzinian, K.~A.~Oganesian, H.~R.~Avakian and E.~De Sanctis,
hep-ph/9908466.
%
%
\bibitem{Boglione:2000jk}
M.~Boglione and P.~J.~Mulders,
Phys.\ Lett.\  {\bf B478} (2000) 114, hep-ph/0001196.
%
%
\bibitem{Diana1}
A.~V.~Efremov, K.~Goeke, M.~V.~Polyakov and D.~Urbano,
Phys.\ Lett.\  {\bf B478} (2000) 94, hep-ph/0001119.
%
%
\bibitem{Avakian:1999rr}
H.~Avakian  [HERMES Collaboration],
Nucl.\ Phys.\ Proc.\ Suppl.\  {\bf 79} (1999) 523;
A.~Airapetian {\it et al.}  [HERMES Collaboration],
Phys.\ Rev.\ Lett.\  {\bf 84} (2000) 4047, hep-ex/9910062.
%
%
\bibitem{Bravar:1999rq}
A.~Bravar  [Spin Muon Collaboration],
Nucl.\ Phys.\ Proc.\ Suppl.\  {\bf 79} (1999) 520.
%
%
\bibitem{JJT97} R.L.\ Jaffe, X.\ Jin, and J. Tang
Phys.\ Rev.\ Lett.\ {\bf 80} (1998) 1166;
Phys.\ Rev.\ {\bf D 57} (1998) 5920.
%
%
\bibitem{Soffer95} J.~Soffer,  Phys.\ Rev.\ Lett.\
{\bf 74} (1995) 1292.
%
%
\bibitem{GJJ95}
G.R.\ Goldstein, R.L.\ Jaffe and X.--D.\ Ji,
Phys.\ Rev.\  {\bf D 52} (1995) 5006.
%
%
\bibitem{Bourrely:1998bx}
C.~Bourrely, J.~Soffer and O.~V.~Teryaev,
Phys.\ Lett.\  {\bf B420} (1998) 375, hep-ph/9710224.
%
%
\bibitem{DPPPW96} D.I.\ Diakonov, V.Yu.\ Petrov, P.V.\ Pobylitsa,
M.V.\ Polyakov and C. Weiss, Nucl.\ Phys.\ {\bf B 480} (1996) 341.
%
\bibitem{DPPPW97} D.I.\ Diakonov, V.Yu.\ Petrov, P.V.\ Pobylitsa,
M.V.\ Polyakov and C. Weiss, 
Phys.\ Rev.\ {\bf D 56} (1997) 4069;
{\it ibid.}\ {\bf D 58} (1998) 038502.
%
\bibitem{DPP88}
D. Diakonov and V. Petrov, Sov.\ Phys.\ JETP
Lett.\ {\bf 43} (1986) 57;  \\
D. Diakonov, V. Petrov and P. Pobylitsa, Nucl.\ Phys.\ 
{\bf B 306} (1988) 809.
%
%
\bibitem{WK98} M. Wakamatsu and T. Kubota, Phys.\ Rev.\ {\bf D 57} 
(1998) 5755; {\it ibid.}\ {\bf D 60}, 034020 (1999).
%
\bibitem{Wakamatsu:2000fd}
M.~Wakamatsu,
hep-ph/0012331.
%
\bibitem{PPGWW98} P.V.\ Pobylitsa, M.V.\ Polyakov, K. Goeke, T. Watabe
and C. Weiss, Phys.\ Rev.\ {\bf D 59} (1999) 034024.
%
\bibitem{Dressler98} For a more detailed discussion of the 
flavor asymmetries, see: B. Dressler, K. Goeke, P.V.\ Pobylitsa, 
M.V.\ Polyakov, T.\ Watabe, and C. Weiss, in: Proceedings of the
11th International Conference on Problems of Quantum
Field Theory, Dubna, Russia, Jul.\ 13--17, 1998, hep-ph/9809487.
%
%
\bibitem{Wakamatsu:2000nj}
M.~Wakamatsu and T.~Watabe,
Phys.\ Rev.\ {\bf D 62} (2000) 054009, hep-ph/9912500.
%
%
\bibitem{Dressler:2000zg}
B.~Dressler, K.~Goeke, M.~V.~Polyakov and C.~Weiss,
Eur.\ Phys.\ J.\  {\bf C14} (2000) 147, hep-ph/9909541.
%
%
\bibitem{Dressler:1999zv}
B.~Dressler, K.~Goeke, M.~V.~Polyakov, P.~Schweitzer, M.~Strikman and C.~Weiss,
hep-ph/9910464. To appear in Eur.~Phys.~J.~C.(2001).
%
%
\bibitem{PP96} P.V.\ Pobylitsa and M.V.\ Polyakov,
Phys.\ Lett.\ {\bf B 389} (1996) 350.
%
%
\bibitem{PPPBGW97} V.Yu.\ Petrov, P.V.\ Pobylitsa, M.V.\ Polyakov,
I. B\"ornig, K. Goeke, and C. Weiss, 
Phys.\ Rev.\ {\bf D 57} (1998) 4325.
%
%
\bibitem{Review} For a review, see: Ch.V.\ Christov {\em et al.},
Prog.\ Part.\ Nucl.\ Phys.\ {\bf 37} (1996) 91.
%
%
\bibitem{Alkofer:1996ph}
R.~Alkofer, H.~Reinhardt and H.~Weigel,
Phys.\ Rept.\  {\bf 265} (1996) 139, hep-ph/9501213.
%
%
\bibitem{Goeke:2000ym}
K.~Goeke, P.~V.~Pobylitsa, M.~V.~Polyakov and D.~Urbano,
Nucl.\ Phys.\ {\bf A680} (2000) 307, hep-ph/0003324.
%
%
\bibitem{Goeke:2000wv}
K.~Goeke, P.~V.~Pobylitsa, M.~V.~Polyakov, P.~Schweitzer and D.~Urbano,
hep-ph/0001272.
%
%
\bibitem{Pobylitsa:2000tt}
P.~V.~Pobylitsa and M.~V.~Polyakov,
Phys.\ Rev.\ {\bf D 62} (2000) 097502, hep-ph/0004094.
%
%
\bibitem{Scopetta:1998qg}
S.~Scopetta and V.~Vento,
Phys.\ Lett.\  {\bf B424} (1998) 25, hep-ph/9706413.
%
%
\bibitem{Gamberg:1998vg}
L.~Gamberg, H.~Reinhardt and H.~Weigel,
Phys.\ Rev.\  {\bf D58} (1998) 054014, hep-ph/9801379.
%
%
\bibitem{Ioffe:1995aa}
B.~L.~Ioffe and A.~Khodjamirian,
Phys.\ Rev.\  {\bf D51} (1995) 3373, hep-ph/9403371.
%
%
\bibitem{Aoki:1997pi}
S.~Aoki, M.~Doui, T.~Hatsuda and Y.~Kuramashi,
Phys.\ Rev.\  {\bf D56} (1997) 433, hep-lat/9608115.
%
%
\bibitem{Kuramashi:1998fi}
Y.~Kuramashi,
Nucl.\ Phys.\  {\bf A629} (1998) 235C, hep-lat/9711015.
%
%
\bibitem{HeJi96}
H.~He and X.~Ji,
Phys.\ Rev.\  {\bf D54} (1996) 6897, hep-ph/9607408.
%
%
\bibitem{Kim96}
H.~Kim, M.~V.~Polyakov and K.~Goeke,
Phys.\ Rev.\  {\bf D53} (1996) 4715, hep-ph/9509283;
Phys.\ Lett.\  {\bf B387} (1996) 577, hep-ph/9604442.
%
%
\bibitem{Artru:1990zv}
X.~Artru and M.~Mekhfi,
Z.\ Phys.\  {\bf C45} (1990) 669.
%
%
\bibitem{Vogelsang:1998ak}
W.~Vogelsang,
Phys.\ Rev.\  {\bf D57} (1998) 1886, hep-ph/9706511.
%
%
\bibitem{Hayashigaki:1997dn}
A.~Hayashigaki, Y.~Kanazawa and Y.~Koike,
Phys.\ Rev.\  {\bf D56} (1997) 7350, hep-ph/9707208.
%
\bibitem{Witten83} E. Witten, Nucl.\ Phys.\ {\bf B 223} (1983) 433.
%
%
\bibitem{DP86}
D. Diakonov and V. Petrov, Nucl.\ Phys.\ {\bf B 245} (1984) 259;
Nucl.\  Phys.\ {\bf B 272} (1986) 457. \\
D. Diakonov and V. Petrov, LNPI preprint LNPI-1153 (1986), published
(in Russian) in: Hadron Matter under Extreme Conditions, Naukova Dumka, 
Kiev (1986), p. 192.
%
%
\bibitem{DE}
D. Diakonov and M. Eides, Sov.\ Phys.\ JETP Lett.\ {\bf 38} (1983) 433; \\
A. Dhar, R. Shankar and S. Wadia, Phys.\ Rev.\ {\bf D 31} (1984) 3256.
%
\bibitem{ANW}
G. Adkins, C. Nappi and E. Witten, Nucl.\ Phys.\ {\bf B 228} (1983) 552.
%
\bibitem{BPW97} J.\ Balla, M.V.\ Polyakov and C. Weiss,
Nucl.\ Phys.\ {\bf B 510} (1997) 327.
%
%
\bibitem{WG97} C. Weiss and K. Goeke, Bochum University preprint
RUB-TPII-12/97, hep-ph/9712447.
%
\bibitem{Fujikawa80} K. Fujikawa, Phys.\ Rev.\ {\bf D 21} (1980) 2848.
%
\bibitem{Blotz95} M.~Prasza\l owicz, A.~Blotz, and 
K.~Goeke, Phys.\ Lett.\ {\bf B 354} (1995) 415.
%
\bibitem{KR84} S.\ Kahana and G.\ Ripka,
Nucl.\ Phys.\ {\bf A 429} (1984) 462.
%
\bibitem{RHIC} G.~Bunce {\it et al.}: ``Polarized protons at RHIC'', in: 
Part.\ World {\bf 3} (1992) 1. \\
The PHENIX/Spin Collaboration: ``Spin Structure Function Physics
with an upgraded PHENIX Muon Spectrometer'', 1994, unpublished, 
available at:\\ 
{\tt http://www.rhic.bnl.gov/phenix} .
%
%
\bibitem{RHICrecent}
G.~Bunce, N.~Saito, J.~Soffer and W.~Vogelsang,
Ann.\ Rev.\ Nucl.\ Part.\ Sci.\ {\bf 50}, 525 (2000), hep-ph/0007218.
%
%
\bibitem{Barone97}
V.~Barone, T.~Calarco and A.~Drago,
Phys.\ Rev.\  {\bf D56} (1997) 527, hep-ph/9702239.
%
%
\bibitem{BourrelySoffer94} C.\ Bourrely and J.\ Soffer,
Nucl.\ Phys.\ {\bf B 423} (1994) 329.
%
\bibitem{Soffer98} J.\ Soffer and J.--M.\ Virey,
Phys.\ Lett.\ {\bf B 314} (1993) 132; 
Nucl.\ Phys.\ {\bf B 509} (1998) 297.
%
\bibitem{Kamal98} B. Kamal, Phys.\ Rev.\ {\bf D 57} (1998) 6663.
%
%
\bibitem{Martin:1999eu}
O.~Martin, A.~Schafer, M.~Stratmann and W.~Vogelsang,
Phys.\ Rev.\  {\bf D60} (1999) 117502, hep-ph/9902250.
%
%
\bibitem{Koepf:1996yh}
W.~Koepf, L.~L.~Frankfurt and M.~Strikman,
Phys.\ Rev.\  {\bf D53} (1996) 2586, hep-ph/9507218.
%
%
\end{thebibliography}
\end{document}